\documentclass[prm,twocolumn,superscriptaddress,amsmath,amssymb,aps,]{revtex4-2}

\usepackage{graphicx}
\usepackage{dcolumn}
\usepackage{bm}
\usepackage{blkarray}
\usepackage{xcolor}
\usepackage{natbib}
\usepackage[pagebackref=false,colorlinks,linkcolor=blue,citecolor=blue,urlcolor=blue]{hyperref}
\usepackage{soul}
\graphicspath{{Figures/}}

\usepackage{amssymb}
\usepackage{amsfonts}
\usepackage{amsmath}
\usepackage{float}
\usepackage{siunitx}
\usepackage{multirow}
\usepackage{hyperref}
\usepackage{xcolor}

\begin{document}

\title{Spin-tunable thermoelectric performance in monolayer  chromium pnictides}

\author{Melania~S.~Muntini}
\email{melania@physics.its.ac.id}
\affiliation{%
Department of Physics, Faculty of Science and Data Analytics, Institut Teknologi Sepuluh Nopember, Surabaya 60111, Indonesia.
}

\author{Edi~Suprayoga}
\email{edi.suprayoga@brin.go.id}
\affiliation{
Research Center for Quantum Physics, National Research and Innovation Agency, Tangerang Selatan 15314, Indonesia.
}

\author{Sasfan~A.~Wella}
\affiliation{
Research Center for Quantum Physics, National Research and Innovation Agency, Tangerang Selatan 15314, Indonesia.
}

\author{Iim~Fatimah}
\affiliation{%
Department of Physics, Faculty of Science and Data Analytics, Institut Teknologi Sepuluh Nopember, Surabaya 60111, Indonesia.
}

\author{Lila~Yuwana}
\affiliation{%
Department of Physics, Faculty of Science and Data Analytics, Institut Teknologi Sepuluh Nopember, Surabaya 60111, Indonesia.
}

\author{Tosawat~Seetawan}
\affiliation{Center of Excellence on Alternative Energy, Research and
Development Institution, Sakon Nakhon Rajabhat University, Sakon
Nakhon 47000, Thailand.
}
\affiliation{Program of Physics, Faculty of Science and Technology,
Sakon Nakhon Rajabhat University, Sakon Nakhon 47000, Thailand.
}

\author{Adam~B.~Cahaya}
\affiliation{
Department of Physics, Faculty of
Mathematics and Natural Sciences, Universitas Indonesia, Depok
16424, Indonesia
}

\author{Ahmad~R.~T.~Nugraha}
\affiliation{
Research Center for Quantum Physics, National Research and Innovation Agency, Tangerang Selatan 15314, Indonesia.
}

\author{Eddwi~H.~Hasdeo}
\email{eddwi.hesky.hasdeo@brin.go.id}
\affiliation{
Research Center for Quantum Physics, National Research and Innovation Agency, Tangerang Selatan 15314, Indonesia.
}%
\affiliation{
Department of Physics and Material Science, 
University of Luxembourg, L-1511 Luxembourg.
}

\date{\today}

\begin{abstract}
Historically, finding two-dimensional (2D) magnets is well known to be a difficult task due to instability against thermal spin fluctuations. Metals are also normally considered poor thermoelectric (TE) materials. Combining intrinsic magnetism in two dimensions with conducting properties, one may expect to get the worst for thermoelectrics. However, we will show this is not always the case. Here, we investigate spin-dependent TE properties of monolayer chromium pnictides (Cr$X$, where $X$ = P, As, Sb, and Bi) using first-principles calculations of electron- and phonon-energy dispersion, along with Boltzmann transport formalism under energy-dependent relaxation time approximation. All the Cr$X$ monolayers are dynamically stable and they also exhibit half metallicity with ferromagnetic ordering. Using the spin-valve setup with antiparallel spin configuration, the half metallicity and ferromagnetism in monolayer Cr$X$ enable manipulation of spin degrees of freedom to tune the TE figure of merit ($ZT$). At optimized chemical potential and operating temperature of $500$~K, the maximum $ZT$ values ($\approx 0.22$, $0.12$, and $0.09$) with the antiparallel spin-valve setup in CrAs, CrSb, and CrBi improve up to almost twice the original values ($ZT \approx 0.12$, $0.08$, and $0.05$) without the spin-valve configuration. Only in CrP, which is the lightest species and less spin-polarized among Cr$X$, the maximum $ZT$ ($\approx 0.34$) without the spin-valve configuration is larger than that ($\approx 0.19$) with the spin-valve one. We also find that, at 500 K, all the Cr$X$ monolayers possess exceptional TE power factors of about $0.02$--$0.08$ W/m.K$^2$, which could be one of the best values among 2D conductors. 
\end{abstract}

\maketitle

\section{\label{sec:intro} Introduction}

\begin{figure}[tb]
    \centering \includegraphics[clip,width=7.5cm]{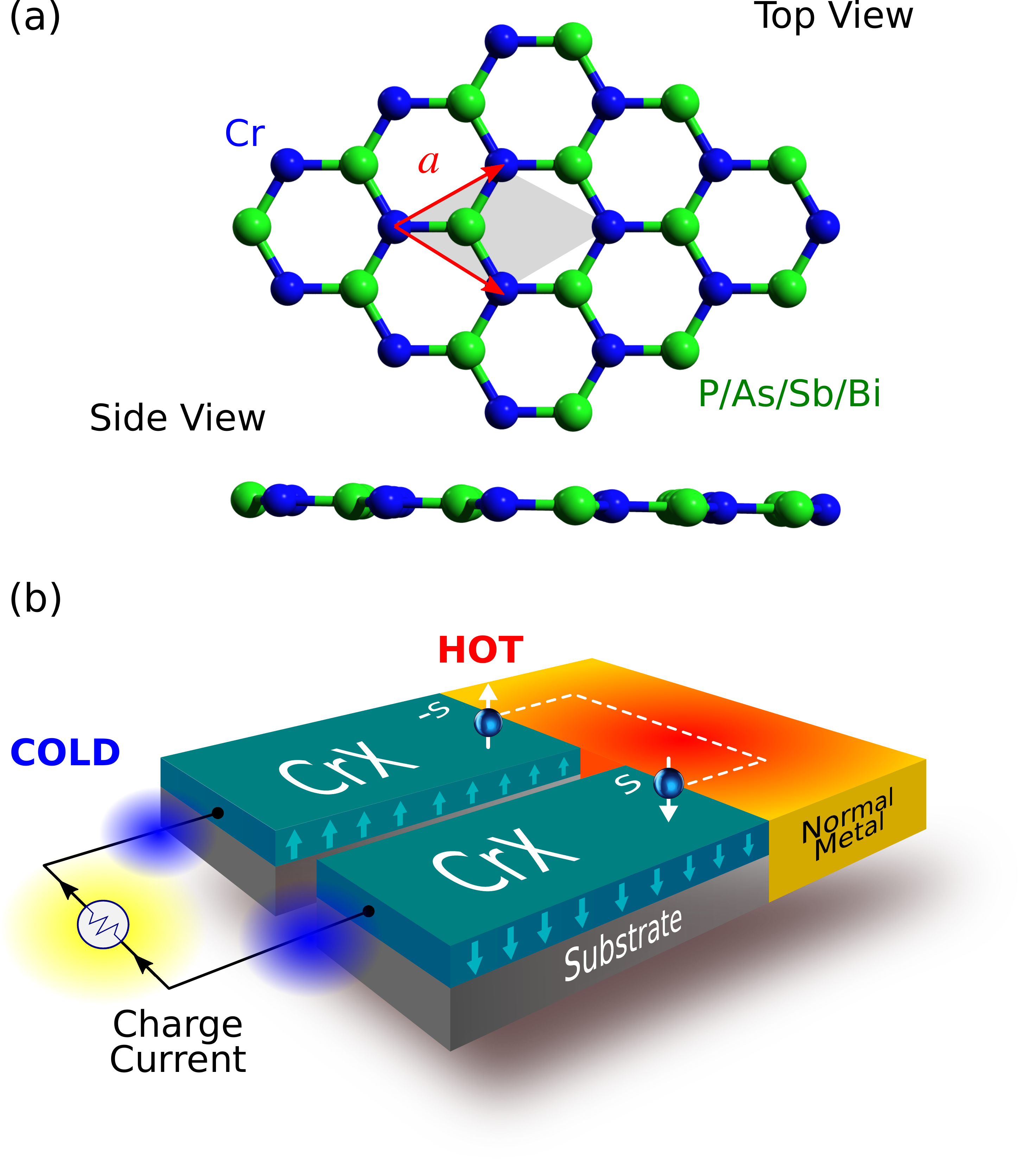}
    \caption{(a) Geometrical structure of 2D chromium pnictides with honeycomb (hexagonal) lattice. (b) Spin-valve device setup with antiparallel spin configuration. }
    \label{fig:structure}
\end{figure}

Thermoelectric (TE) materials convert heat directly into electricity so that they could be beneficial for powering electronic devices in some situations where heat loss is abundant~\cite{goldsmid2010}, such as from automotive engines, high-performance computers, or even human bodies. For most of the applications at moderate temperature (around 500 K), the cost-effective power generation of the TE materials requires a high power factor (PF) on the order of at least $10^{-3}$ W/m.K$^2$~\cite{liu2016}. Besides PF, there is also an efficiency-related quantity known as the dimensionless figure of merit, $ZT$, which one can calculate by the following formula~\cite{goldsmid2010}:
\begin{equation}
ZT = \frac{\mathrm{PF}}{\kappa} T = \frac{S^2 \sigma}{\kappa_e + \kappa_{ph}}  T,
\end{equation}
where $S$ is the Seebeck coefficient, $\sigma$ is the electrical conductivity, $T$ is the operating temperature, and $\kappa$ is the total thermal conductivity (sum of the electronic contribution $\kappa_e$ and lattice (or phonon) contribution $\kappa_{ph}$). Therefore, in conventional thermoelectrics, to obtain as large $ZT$ as possible, we need simultaneously high $S$, high $\sigma$, and low $\kappa$. This condition is, unfortunately, difficult to achieve by nature~\cite{minnich2009,mona2012,heremans2013}.  
For example, in normal metals, $\sigma$ and electronic part of $\kappa$ (denoted by $\kappa_e$) depend on each other through the Wiedemann-Franz law~\cite{wflaw}, which prohibits very high $\sigma$ and very low $\kappa$ to emerge simultaneously.
Furthermore, $S$ ($\sigma$) in metals or other conducting materials is lower (higher) than $S$ ($\sigma$) in semiconductors or insulators~\cite{shakouri2011}. In this regard, much of thermoelectrics research in the past decades has often focused on semiconductors to find optimal values of the TE transport coefficients ($S$, $\sigma$, and $\kappa$) that can give $ZT > 1$~\cite{dresselhaus2007}. On the other hand, most conductors usually only possess $ZT < 0.1$~\cite{markov2018}. 

Recent advances in two-dimensional (2D) materials have also opened up an opportunity for exploring high-performance thermoelectrics in low dimensions~\cite{markov2019,kanahasi2020,li2020}. For example, black phosphorus~\cite{fei2014,bolin2015,iwasa2016}, transition metal dichalcogenides~\cite{jin2015,yoshida2016}, and group-III chalcogenides~\cite{wickramaratne2015,zeng2018} in the family of 2D semiconductors have been found to exhibit better TE performance than their bulk counterparts. Pioneering studies by Dresselhaus' group have pointed out that the quantum confinement effect plays a crucial role in improving PF or ZT in low dimensions~\cite{hicks93,hicks96}. In particular, when the confinement length is less than the thermal de Broglie wavelength of the material under consideration, the PF enhancement is theoretically guaranteed to be achieved~\cite{hung2016-prl}. However, the theory assumed the material does not have intrinsic magnetism and is not spin-polarized. Therefore, to obtain high-performance thermoelectrics in 2D materials with intrinsic magnetism, it seems that one needs to consider another enhancement technique beyond quantum confinement and find the materials that can utilize the mechanism.

Before looking for suitable 2D magnets for thermoelectrics, one should note that the difficulty of obtaining long-range magnetic order in two dimensions is a long-standing problem that prevents most 2D materials from possessing intrinsic magnetism. According to the Mermin-Wagner theorem, the long-range magnetic order in isotropic 2D magnets at finite temperatures is unstable against thermal spin fluctuations~\cite{merminwagner}. However, by the presence of magnetic anisotropy, some materials such as CrI$_3$ and Cr$_2$Ge$_2$Te$_6$ can emerge as 2D magnets with the magnetic order depending on the number of layers~\cite{gong2017,huang2017}. Analyzing the structural simplicity of CrI$_3$, one may expect the other 2D Cr-based honeycomb structures [Fig.~\ref{fig:structure}(a)] may also lead to intrinsic magnetism. Indeed, recent studies on monolayer chromium pnictides (Cr$X$, where $X$ = P, As, Sb, and Bi) suggested that monolayer Cr$X$ can be another 2D magnet, yet half-metallic. While monolayer CrI$_3$ as a \emph{semiconductor} is potentially a good TE material~\cite{gao2020,sheng2020}, we do not know whether monolayer Cr$X$ can perform similarly to CrI$_3$. Note that CrI$_3$ also exhibit spin-dependent TE properties, with the $ZT$ value, at 500 K, of about $0.26$, which is high enough for the family of 2D materials (c.f. monolayer MoS$_2$ with $ZT \sim 0.11$~\cite{jin2015} and InSe with $ZT \sim 0.5$~\cite{nguyen2017} at the same temperature). However, the constant relaxation time approximation used in Refs.~\cite{gao2020} and~\cite{sheng2020} might not be accurate enough for the $ZT$ calculation. Moreover, although the TE transport coefficients in CrI$_3$ are spin-dependent, there is no clear advantage of the spin polarization for the $ZT$ enhancement in CrI$_3$ because one cannot utilize the spin degrees of freedom in semiconductors using the spin-valve TE device~\cite{Cahaya}. By contrast, half metallicity in monolayer Cr$X$ may allow manipulation of spin degrees of freedom to enhance the $ZT$.

In this work, we will show our simulation suggesting that the $ZT$ values in monolayer Cr$X$ can increase using the spin-valve setup with antiparallel configuration [Fig.~\ref{fig:structure}(b)].  Since the majority carriers (spin-up states) are metallic and the minority ones (spin-down states) are semiconducting, the PF values in monolayer Cr$X$ are also exceptionally high for a 2D material because the minority carriers contribute to a high Seebeck coefficient, while the majority carriers contribute to high electrical conductivity. We perform the calculations for TE transport coefficients within the linearized Boltzmann transport theory and energy-dependent relaxation time approximation, with electronic energy and phonon dispersion relations obtained from first-principles density functional theory (DFT), as outlined in the next section.

\section{\label{sec:comp} Computational Methods}

All optimized geometrical structures, electronic properties, and phonon dispersion relations of Cr$X$ monolayers are calculated with DFT as implemented in the {\sc{Quantum ESPRESSO}} code~\cite{QE}. We employ the optimized norm-conserving Vanderbilt (ONCV) pseudopotentials~\cite{hamann2013, schlipf2015} to describe the interaction between electrons and ions. We use the Perdew-Burke-Ernzerhof functional~\cite{pbe1996} under generalized gradient approximation to describe exchange-correlation energy and potential. The wave functions are expanded in plane-wave basis sets with the cutoff energy as high as $\sim$820 eV. We sample the Brillouin zone using dense $32 \times 32 \times 1$ and $64 \times 64 \times 1$ Monkhorst-Pack (MP) grids for calculations of optimized geometry and density of states, respectively. All structures are relaxed until the maximum Hellmann-Feynman force per atom is less than $0.26$~meV/\AA. We set a vacuum layer to 30 \AA{} to avoid the interlayer interactions due to the lattice periodicity. We then obtain the optimized lattice constants $a$ [see Fig.~\ref{fig:structure}(a)] of $3.88$, $3.94$, $4.39$, and $4.58$~\AA{} for CrP, CrAs, CrSb, and CrBi, respectively, in good agreement with available data in the literature~\cite{Mogulkoc2020,Mogulkoc2021}. With the optimized geometry data, we can calculate the ground-state electronic structures of monolayer Cr$X$. 

We use DFT+$U$ improvement of accuracy, especially for Cr, to include the effects of strong electronic correlations.  Following a prior theoretical study~\cite{Mogulkoc2020}, $U=3$ eV can be employed for Cr atoms with the spin polarization taken into account for all the systems. 
The initial magnetic moment of each Cr$X$ is set to three times Bohr magneton ($\mu_B$). After relaxation, the resulting magnetic moments of CrP, CrAs, CrSb, and CrBi monolayers are $2.97\mu_B$, $3.00 \mu_B$, $3.00 \mu_B$, and $3.00\mu_B$, respectively, which agree well with the reported value of $3\mu_B$ in the literature~\cite{Mogulkoc2020,Mogulkoc2021}.
For the spin-unpolarized band structures, readers can refer to Supplementary Material Fig.~S1~\cite{SM}. 
Although Cr$X$ monolayers here posssess non-zero magnetic moments, the DFT+$U$ method is already sufficient and consistent with the DFT+$U$+$J$ method because the $J$ parameter for these cases was found to be small enough, around $10$~meV~\cite{Mogulkoc2020,Mogulkoc2021}.
The dynamical stability of Cr$X$ is also confirmed by calculating the phonon dispersion [Fig.~S2 in the Supplementary Material] with the dynamical matrix evaluated on the $4\times4\times1$ MP grid of \textbf{q}-points~\cite{SM}.
Note that we do not consider van der Waals interaction in this study because it will not change the band structure significantly; it only gives a minor change in the lattice constants~\cite{Mogulkoc2020,Mogulkoc2021}.

Having complete information of electronic structures, we calculate the TE transport coefficients using {\sc{BoltzTraP2}} code~\cite{boltztrap2} within linearized Boltzmann transport theory and energy-dependent relaxation time approximation. The moment of the generalized transport coefficient with spin index $j=\uparrow$ and $\downarrow$ is given by,
\begin{equation}
	\mathcal{L}^{(\alpha)}_j = e^2 \sum_{n,\textbf{k}} \tau_{n,\textbf{k}}v^2_{n,\textbf{k}}(\varepsilon_{n,\textbf{k}}-\mu)^\alpha \left(-\frac{\partial f_{n,\textbf{k}}}{\partial \varepsilon_{n,\textbf{k}}} \right).
	\label{eq.2}
\end{equation}
This kernel is used to calculate the electrical conductivity, the Seebeck coefficient, and the electron thermal conductivity as follows:
\begin{equation}
	\sigma_j = \mathcal{L}^{(0)},\quad
	S_j = \frac{1}{eT} \frac{\mathcal{L}^{(1)}}{\mathcal{L}^{(0)}},\quad
	\kappa_{e,j} = \frac{1}{e^2T} \left[\mathcal{L}^{(2)} - \frac{\mathcal{L}^{(1)2}}{\mathcal{L}^{(0)}} \right],
	\label{eq.3}
\end{equation}
where $e$ is the unit electric charge, $\mu$ is the chemical potential, $\varepsilon_{n,\textbf{k}}$ is the energy of $n$th band at wave vector $\textbf{k}$, $\tau_{n,\textbf{k}}$ is the electronic relaxation time, $v_{n,\textbf{k}}$ is the electronic group velocity, and $f_{n,\textbf{k}}$ is the Fermi-Dirac distribution function.  To fairly treat $\sigma$ and $\kappa_e$ of 2D systems, we multiply the output $\sigma$ and $\kappa_e$ from {\sc{BoltzTraP2}} with the thickness of the simulation box over the conventional width of the monolayer.  The total contributions from both spins read as:
\begin{equation}
    \sigma = \sigma_\uparrow + \sigma_\downarrow,\quad S = \frac{\sigma_\uparrow S_\uparrow+\sigma_\downarrow S_\downarrow}{\sigma},\quad \kappa_e = \kappa_{e,\uparrow}+\kappa_{e,\downarrow}.
    \label{eq:total}
\end{equation}

We consider the electron-phonon scattering as the most dominant scattering contribution to the electronic relaxation time, which can be defined as $\tau_{n,\textbf{k}}=\hbar/(2~\mathrm{Im}\sum_{n,\textbf{k}})$, where $\hbar$ is the reduced Planck constant and $\mathrm{Im}\sum_{n,\textbf{k}}$ is the imaginary part of the electronic self energy.  Using the {\sc{EPW}} code \cite{epw}, we compute the self energy using the electronic energy and the phonon dispersion on relatively coarse $16\times16\times1$ \textbf{k}-point and $8\times8\times1$ \textbf{q}-point grids, respectively.  To obtain a finer result, we fit $1/\tau_{n,\textbf{k}}$ from {\sc{EPW}} with an energy-dependent $\tau$ model:
\begin{equation}
    1/\tau(\varepsilon)=\mathcal{C}\cdot {\mathrm{DOS}}(\varepsilon)+ 1/\tau_0,
\end{equation}
where $\mathcal{C}$ is a DOS-dependent fitting parameter and $\tau_0$ is the relaxation time constant that is chosen to overcome the absence of $\mathrm{DOS}$ in the band gap area in the spin-down band.  
The resulting $1/\tau(\varepsilon)$ model for each Cr$X$ species is given in Fig.~S3~\cite{SM}.
Note that although the CrX family are ferromagnetic materials, we do not consider contribution of magnetic spin excitations (known as magnons) to the thermoelectric properties. The magnon effect in heterostructures [like the system depicted in Fig.~\ref{fig:structure}(b)] is suppressed because the magnons cannot escape the ferromagnets~\cite{PhysRevB.79.174426}.

Finally, to fully obtain $ZT$, we calculate the phonon thermal conductivity $\kappa_{ph}$ from the linearized phonon Boltzmann transport equation for phonons within the single-mode relaxation-time method, as implemented in the {\sc{Phono3py}} code~\cite{phono3py1,phono3py2}. The second-order harmonic and third-order anharmonic interatomic force constants are computed using a $2\times2\times1$ supercell with the finite displacement method. To obtain a converged result, we use $32\times32\times1$ \textbf{q}-point mesh to sample reciprocal space of the primitive cells of all Cr$X$ monolayers.

\begin{figure*}[t]
    \centering \includegraphics[clip,width=15cm]{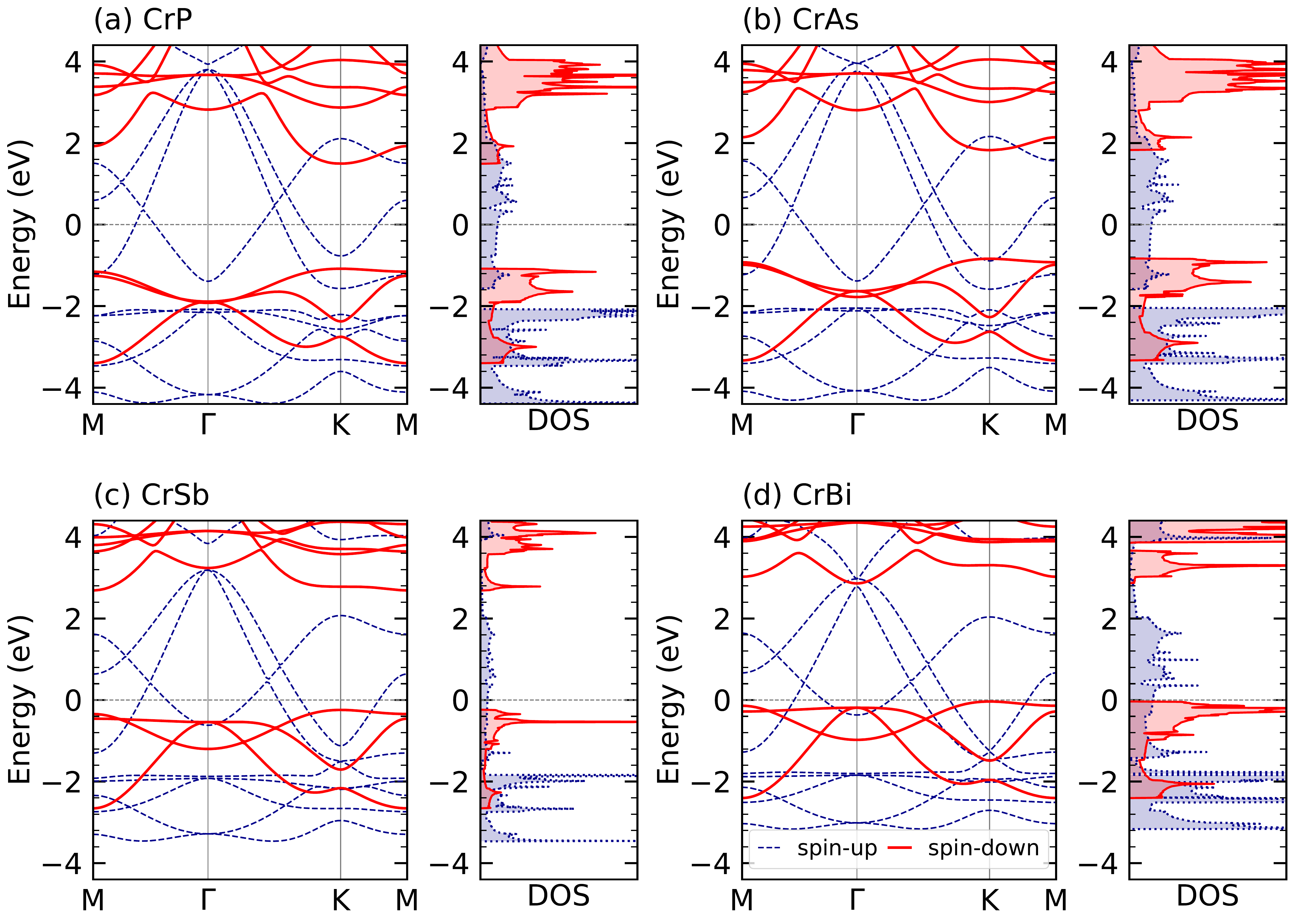}
    \caption{%
    Electronic structure (energy dispersion and density of states) of chromium pnictides: (a) CrP, (b) CrAs, (c) CrSb, (d) CrBi. The blue dashed (red solid) lines refer to spin-up (-down) states.
    %
    }
    \label{fig:bands}
\end{figure*}

\section{Results and discussion}

The Cr$X$ monolayers have a honeycomb structure without buckling along an axis normal to the surface, similar to graphene [see Fig.~\ref{fig:structure}]. Their electronic structures, on the other hand, do not correspond to graphene mostly because of the presence of $d$ orbitals in Cr. The $d$ orbitals play a significant role to lift the spin degeneracy.
In Fig.~\ref{fig:bands}, we show the electronic structures of (a)~CrP, (b)~CrAs, (c)~CrSb, and (d)~CrBi, which are all consistent with previous first-principles results~\cite{Mogulkoc2020,Mogulkoc2021}.  In particular, one can compare the good agreement of Fig.~\ref{fig:bands} in this study with Fig. 4 of Ref.~\cite{Mogulkoc2020}.
The dashed (solid) line refers to spin-up (down) states.  Spin-up states are metallic while spin-down states possess band gaps displaying the half-metallic phase. At the Fermi level, chromium pnictides are ferromagnet. These electronic structures are distinct from monolayer chromium iodide ($\mathrm{CrI_3}$ which shows ferromagnetic insulator~\cite{gong2017}. As atomic number increases from CrP to CrBi, both bandwidth and half-metallic band gap decrease.  We will show later how these behaviors will tune the TE transport. 

\begin{figure*}[t]
    \centering \includegraphics[clip,width=15cm]{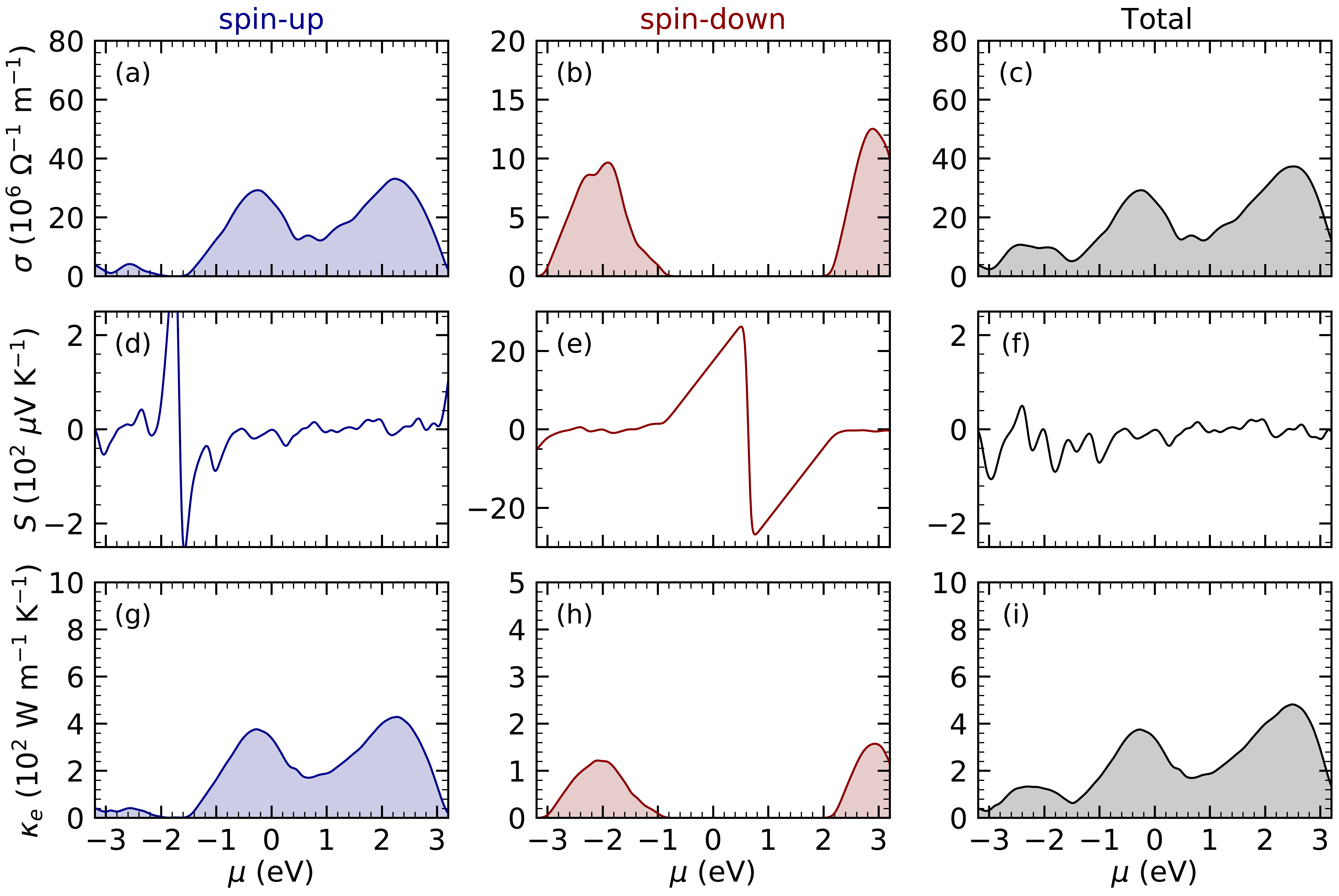}
    \caption{Thermoelectric quantities of CrP: (a-c) Electrical conductivity, (d-f) Seebeck coefficient, and (g-i) electron thermal conductivity of spin-up, spin-down, and total, respectively. }
    \label{fig:TE-CrP}
\end{figure*}

In Fig.~\ref{fig:TE-CrP}, we plot the TE transport coefficients as a function of chemical potential for CrP as a representative of chromium pnictides at $T=500~\rm K$.  Meanwhile, the TE transport coefficients of CrAs, CrSb, and CrBi are given in Figs.~S4-S6~\cite{SM}. As expected, the spin-up conductivity dominates over the spin-down conductivity as depicted in Figs.~\ref{fig:TE-CrP}(a) and ~\ref{fig:TE-CrP}(b). Therefore, the total conductivity $\sigma$ pretty much resembles the spin-up contribution [Fig.~\ref{fig:TE-CrP}(c)]. On the contrary, the Seebeck effect is dominated by the spin-down states because of the presence of the band gap. According to the Mott formula, the Seebeck coefficient is proportional to $-\sigma'(\mu)/\sigma(\mu)$; thus it will be large if the band gap exists. For the spin-down states, the Seebeck coefficient reaches $2000~\mathrm{\mu V/K}$. Normally, the Seebeck coefficient is proportional to the band gap $\Delta$ in the non-degenerate limit ($k_BT\ll\Delta $). For the spin-up states, the Seebeck coefficient is not entirely zero, typically on the order of fundamental entropy per charge ($k_B/e=87~ \mathrm{\mu V/K}$).

The total Seebeck coefficients $S$ are weighted by their conductivity so that the results mostly follow the shape of the spin-up states that indicates dominant contributions from majority carrier at the Fermi level. Despite the small $S$ values, their oscillations around zero will play an important role later when spin-dependent transport is considered. In Figs.~\ref{fig:TE-CrP}(g) and~\ref{fig:TE-CrP}(i), the spin-up conductivity and total electron thermal conductivity $\kappa_e$ resemble the shape of $\sigma$, consistent with the Wiedemann-Franz law. In fact, the Lorenz number $\kappa_e/\sigma T$ is $2.7\times 10^{-8}\ \rm W.\Omega/K^{2}$, similar to those of ordinary metals. 

\begin{figure}[tb]
    \includegraphics[clip,width=7.5cm]{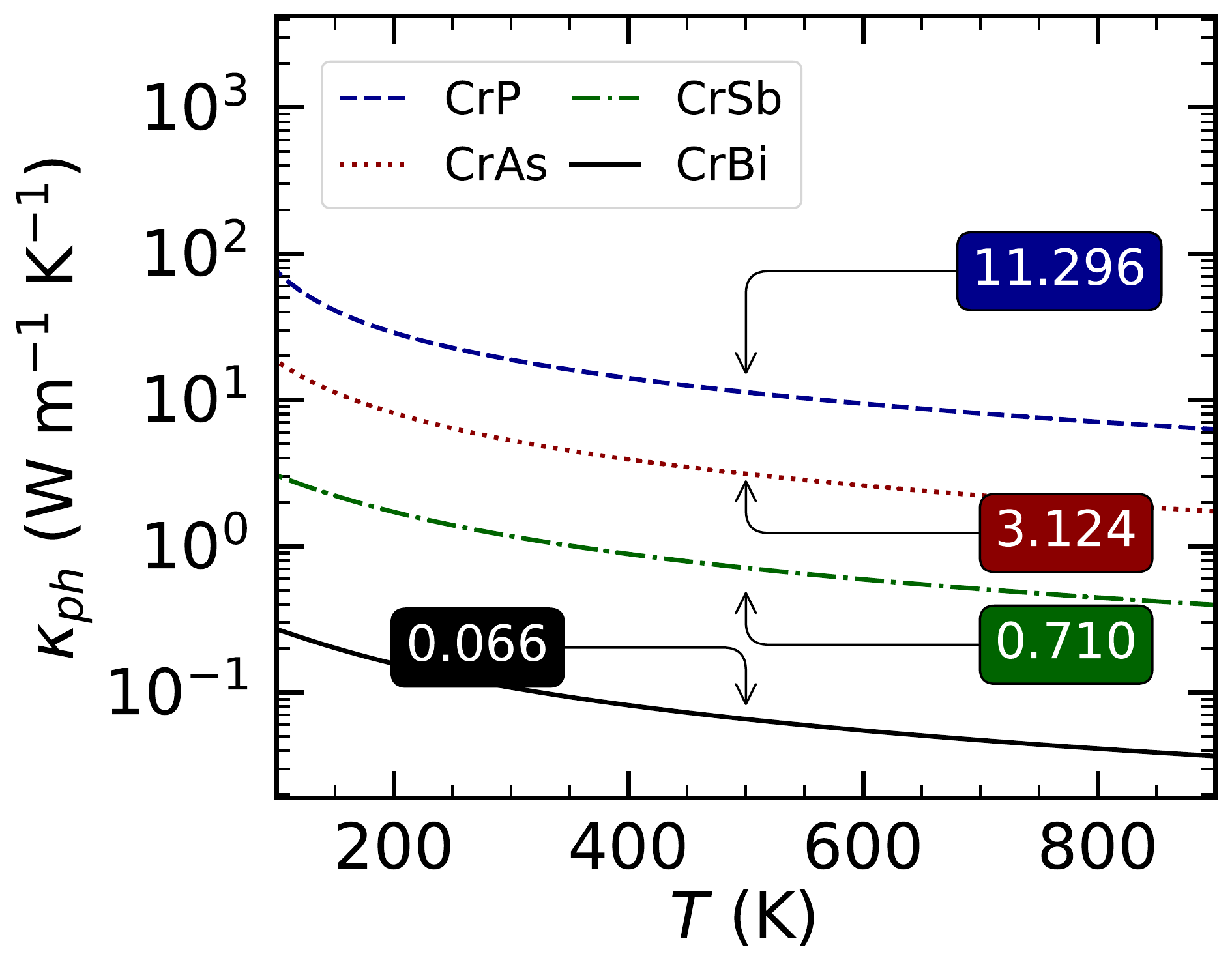}
    \caption{Phonon thermal conductivity $\kappa_{ph}$ as a function of temperature for different Cr$X$ species. Indicated by boxes are $\kappa_{ph}$ values at 500 K.}
    \label{fig:Kappaph}
\end{figure}

\begin{figure*}[t]
  \centering \includegraphics[width=17cm]{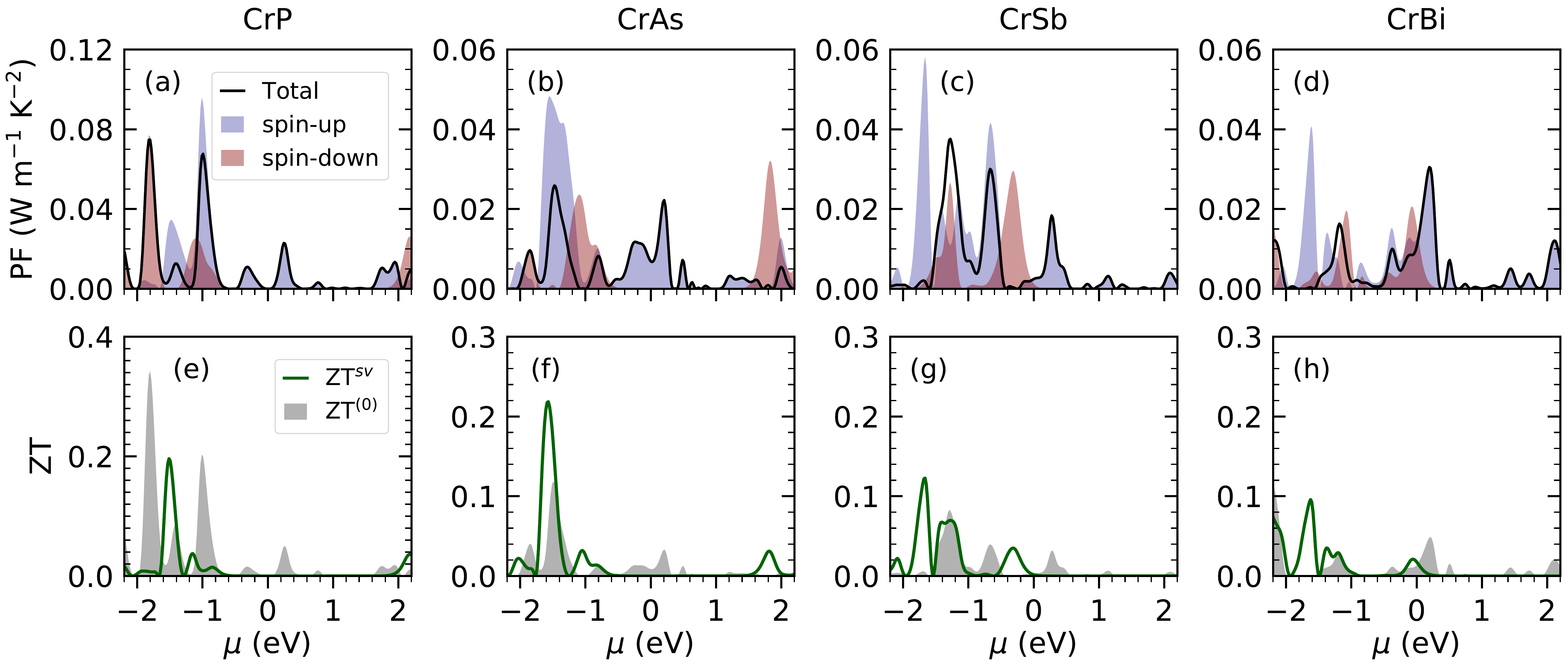}
  \caption{(a-d) Power factors and (e-f) figures of merit of CrP, CrAs, CrSb, and CrBi, respectively, as a function of chemical potential. In the PF plots, we show the contribution of spin-up states, spin-down states, and their total contribution to the PF values. In the ZT plots, with $T = 500$~K, we show comparison of total $ZT^{(0)}$ [Eq.~\eqref{eq:ZT}] and $ZT^{\rm sv}$ that includes the spin-valve effect. }
\label{fig:PF-ZT}
\end{figure*}

We show phonon thermal conductivity $\kappa_{ph}$ of all Cr$X$ monolayers as a function of $T$ in Fig.~\ref{fig:Kappaph}
with the assumption that the magnon effect is neglected and that the materials are kept ferromagnetic. The Curie temperatures of CrX monolayers are within $550$--$700$~K~\cite{Mogulkoc2020}, above which the ferromagnetic to paramagnetic phase transition and short-range magnetic interaction may occur.  Therefore, to avoid the issue of phase transition and short-range magnetic interaction, we can focus on the practicality of our system below $550$~K. For example, as indicated by the arrow and small box in Fig.~\ref{fig:Kappaph}, at $T=500$~K, 
$\kappa_{ph}$ of CrP is smaller by almost two-order of magnitudes than $\kappa_e$. This comparison generally holds for the other chromium pnictides. This result suggests that sole electronic transport calculation for the TE properties is sufficient as the phonon contribution to $\kappa$ is minimum. As the atom gets heavier from CrP to CrBi, $\kappa_{ph}$ decreases following the Debye-Callaway model. 

We turn our analysis on the PF and $ZT$ of Cr$X$ as shown in Fig.~\ref{fig:PF-ZT}. The total PF is dominated by the spin-up states as expected. With the increase of atomic weight, maximum PF tends to decrease but PF at $\mu=0$ increases. CrP has the highest PF because it has the largest Fermi velocity while CrBi has a high PF at $\mu=0$ because it has an additional contribution from the spin-down state at that energy.  It is worth mentioning that the PF of Cr$X$ (about $0.02$--$0.08$ W/m.K$^2$) could be one of the best values among 2D conductors, comparable to the extraordinary PF record in Mg$_3$Bi$_2$-based materials~\cite{mao2019high}.  The total figure of merit $ZT^{(0)}$, given by
\begin{align}
    ZT^{(0)}=\lim_{\sigma_\uparrow\gg \sigma_\downarrow}\frac{\left(\sigma_\uparrow S_\uparrow+\sigma_\downarrow S_\downarrow\right)^2T}{\left(\sigma_\uparrow+\sigma_\downarrow\right)\kappa}= \frac{\sigma_\uparrow S_\uparrow^2T}{\kappa}
    \label{eq:ZT}
\end{align}
is also dominated by the spin-up states with a value of about $10^{-3}$ at small $\mu$ and reaching maximum value of $\sim 0.1$ at a large hole doping [see Figs.~\ref{fig:PF-ZT}(e--h) gray area].  The figure of merit $ZT^{(0)}$ behaves similarly with the PF trend, in which CrP possesses the largest one.  The figure of merit also decreases as the atomic weight increases. 

So far, we have seen Cr$X$ as ordinary metals with all TE properties are dominated by the majority carrier (spin-up states).  The spin-down states, on the other hand, are gapped, and they possess a high Seebeck coefficient.  It is possible to incorporate the spin-down contribution in this system to enhance the $ZT$ values using a special configuration~\cite{Cahaya, PhysRevLett.99.066603}. By inserting a non-magnetic metal spacer between two ferromagnetic materials and shifting their magnetization orientation by $\theta$, voltage generation can increase due to finite spin accumulation~\cite{Cahaya, PhysRevB.79.174426}
\begin{align}
    \frac{S(\theta)}{S}= \frac{1+(1-P'P)\zeta(\theta)}{1+(1-P^2)\zeta(\theta)},
    \label{eq:stheta}
\end{align}
where $P=\left(\sigma_\uparrow -\sigma_\downarrow \right)/\left(\sigma_\uparrow+ \sigma_\downarrow\right)$ is the spin polarization of the conductivity, $P'=\left(\sigma_\uparrow S_\uparrow-\sigma_\downarrow S_\downarrow\right)/\left(\sigma_\uparrow S_\uparrow+\sigma_\downarrow S_\downarrow\right)$ is the spin polarization of the Seebeck current and
\begin{align}
    \zeta(\theta)\propto\frac{V_s(\theta)}{J_s(\theta)}\propto \frac{1-\cos \theta}{1+\cos\theta} = \tan^2\frac{\theta}{2}
\end{align}
that depends on the ratio of spin accumulation $V_s(\theta)$ and spin current $J_s(\theta)$ at the spacer.  Note that since Cr$X$ are half metals, their value of $P$ is around $1$ across the half-metallic gap while $P'$ can be large when the sign of $S_\uparrow$ and $S_\downarrow$ are opposite to each other.  This $P'$ value can be large when $S_\uparrow$ oscillates around zero as shown in Fig.~\ref{fig:TE-CrP}(d).  However, the large value of $P'$ also means that the total Seebeck coefficient is small [see Eq.~\eqref{eq:total}].  

Cahaya et al. introduced antiparallel spin configuration ($\theta=\pi$) as shown in Fig.~\ref{fig:structure}(b) that maximizes voltage and temperature gradients~\cite{Cahaya}. This configuration utilizes a $p$-$n$ junction (a junction with the opposite value of Seebeck coefficient at two ends) and maximizes spin accumulation at its spacer. A nonmagnetic metal that connects $p$- and $n$-types Cr$X$ materials acts as a spin accumulator that ensures spin current is minimized. Using the antiparallel configuration, we show spin-valve effect on the figure of merit $ZT^{\rm sv}$ by using $S(\pi)$ in Eq.~\eqref{eq:stheta} to Eq.~\eqref{eq:ZT}.  Solid green lines in Figs.~\ref{fig:PF-ZT}(e--h) show that $ZT^{(0)}$ can be enhanced by a factor of two with this configuration. 

\begin{figure}[tb]
  \includegraphics[width=8.5cm]{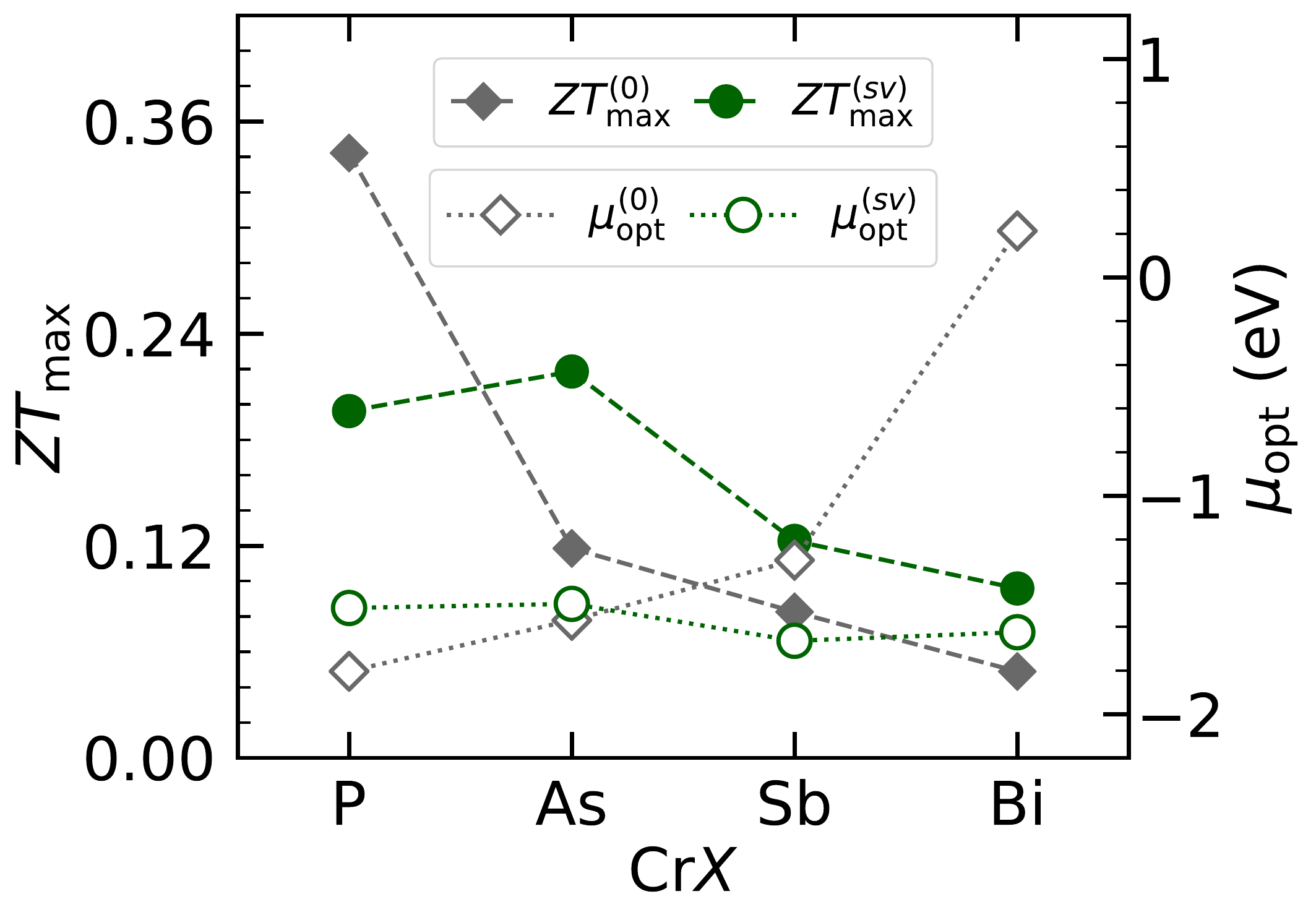}
  \caption{Maximum $ZT^{(0)}$ and $ZT^{\rm sv}$ (left axis) and the corresponding optimal chemical potential $\mu_{\rm opt}$ (right axis) for each Cr$X$ species with $T = 500$~K.}
\label{fig:Summary}
\end{figure}

We summarize the maximum $ZT$ ($ZT_{\rm max}$) values along with their optimum chemical potential for all the Cr$X$ monolayers in Fig.~\ref{fig:Summary}.  At the optimized chemical potential $\mu_{\rm opt}$ and operating temperature of $T=500$~K, the $ZT_{\rm max}$ values ($\approx 0.22$, $0.12$, and $0.09$) with the antiparallel spin-valve setup in CrAs, CrSb, and CrBi improve up to almost twice the original values ($ZT^{(0)} \approx 0.12$, $0.08$, and $0.05$) without the spin-valve configuration.  Only in CrP, which is the lightest species and less spin-polarized among Cr$X$, the $ZT_{\rm max}$ value ($\approx 0.34$) without the spin-valve configuration is larger than that ($\approx 0.19$) with the spin-valve one. Optimized $\mu$ occurs at large hole doping around $-1$ eV to $-2$ eV with only exception for CrBi, whose $ZT^{(0)}_{\rm max}$ occurs near $\mu=0$.  Note that both $ZT^{(0)}$ and $ZT^{\rm sv}$ have weak temperature dependence because of strong resemblance to the Wiedemann-Franz law. 

One may question that the requirement of large doping to obtain the $ZT$ value ($< 0.4$ for the Cr$X$ monolayers) in this work is still disappointing for practical interest of using Cr$X$ in real TE devices.  Nevertheless, we have shown that, by using first-principles calculations with almost the best level of approximation in thermoelectrics theory, the enhancement of $ZT$ due to the half metallicity and spin polarization is truly possible.  As for the practical interest, in addition to $ZT$, we can also talk about the PF, which is useful when the heat input (source) is abundant. For example, the PF value of CrBi in this work can reach $0.03$ W/m.K$^2$, which is an excellent value for a 2D material, at chemical potential near zero, thus it does not require large doping. Further design of the metallic 2D magnets to obtain larger $ZT$ at low chemical potential shall be an interesting problem of further studies.

\section{Conclusions}
Using monolayer Cr$X$ as a prototype, we have proven that the TE performance of half-metallic 2D ferromagnets can improve due to the spin-dependent TE transport coefficients.
The synergistic effects of different spin degrees of freedom are possible through the antiparallel spin-valve configuration, which leads to the $ZT$ enhancement up to almost twice the original $ZT$ values. The origin of the $ZT$ enhancement is that the majority carriers with spin-up states contribute to high electrical conductivity. On the other hand, the minority carriers with spin-down states contribute to a high Seebeck coefficient. The same principle also leads to the exceptional PF in monolayer Cr$X$ within $0.02$--$0.08$ W/m.K$^2$. This work paves the way toward selecting suitable 2D ferromagnets for TE applications.

\begin{acknowledgments}

M.S.M. acknowledges financial support from the research and community service center, Institut  Teknologi Sepuluh Nopember, Indonesia.  E.H.H. acknowledges financial support by Fonds National de la Recherche Luxembourg, Core 14764976 and 15752388. E.S., S.A.W., A.R.T.N, and E.H.H. performed numerical calculations with Mahameru High-Performance Computing (HPC) facilities provided by BRIN and HPC facilities of the University of Luxembourg (see http://hpc.uni.lu). This work was supported by e-ASIA JRP.

\end{acknowledgments}

\bibliographystyle{apsrev4-2}
%


\widetext
\clearpage

\setcounter{equation}{0}
\setcounter{figure}{0}
\setcounter{table}{0}
\setcounter{page}{1}
\setcounter{section}{0}
\renewcommand{\thepage}{S\arabic{page}}  
\renewcommand{\thesection}{S\arabic{section}}   
\renewcommand{\thetable}{S\arabic{table}}   
\renewcommand{\thefigure}{S\arabic{figure}}
\renewcommand{\theequation}{S\arabic{equation}}
\renewcommand{\bibnumfmt}[1]{[S#1]}
\renewcommand{\citenumfont}[1]{S#1}

\begin{center}
\textbf{\large Supplementary Material for:\\
Spin-tunable thermoelectric performance in monolayer 
chromium pnictides}\\[.5cm]

Melania~S.~Muntini,$^{1,*}$ Edi~Suprayoga,$^{2, \dagger}$ Sasfan~A.~Wella,$^{2}$ Iim~Fatimah,$^{1}$ Lila~Yuwana,$^{1}$ Tosawat~Seetawan,$^{3,4}$ Adam~B.~Cahaya,$^{5}$ Ahmad~R.~T.~Nugraha,$^{2}$ Eddwi~H.~Hasdeo$^{2,6,\ddagger}$\\[.4cm]

{\itshape ${}^1$Department of Physics, Faculty of Science and
Data Analytics, Institut Teknologi Sepuluh Nopember, 
Surabaya 60111, Indonesia\\
${}^2$Research Center for Physics, National Research and Innovation Agency (BRIN), South Tangerang, 15314, Indonesia\\
${}^3$Center of Excellence on Alternative Energy, Research and
  Development Institution, Sakon Nakhon Rajabhat University, Sakon
  Nakhon 47000, Thailand\\
${}^4$Program of Physics, Faculty of Science and Technology,
  Sakon Nakhon Rajabhat University, Sakon Nakhon 47000, Thailand\\
${}^5$Department of Physics, Faculty of
  Mathematics and Natural Sciences, Universitas Indonesia, Depok
  16424, Indonesia\\
${}^6$Department of Physics and Materials Science, University of Luxembourg, Luxembourg 1511, Luxembourg}
${}^*$Electronic address: melania@physics.its.ac.id\\
${}^\dagger$Electronic address: edis008@brin.go.id\\
${}^\ddagger$Electronic address: eddw001@brin.go.id\\
(Dated: \today)\\[1cm]
\end{center}

\section{Electronic band structure without spin polarization}

\begin{figure*}[h!]
  \centering \includegraphics[width=0.85\textwidth]{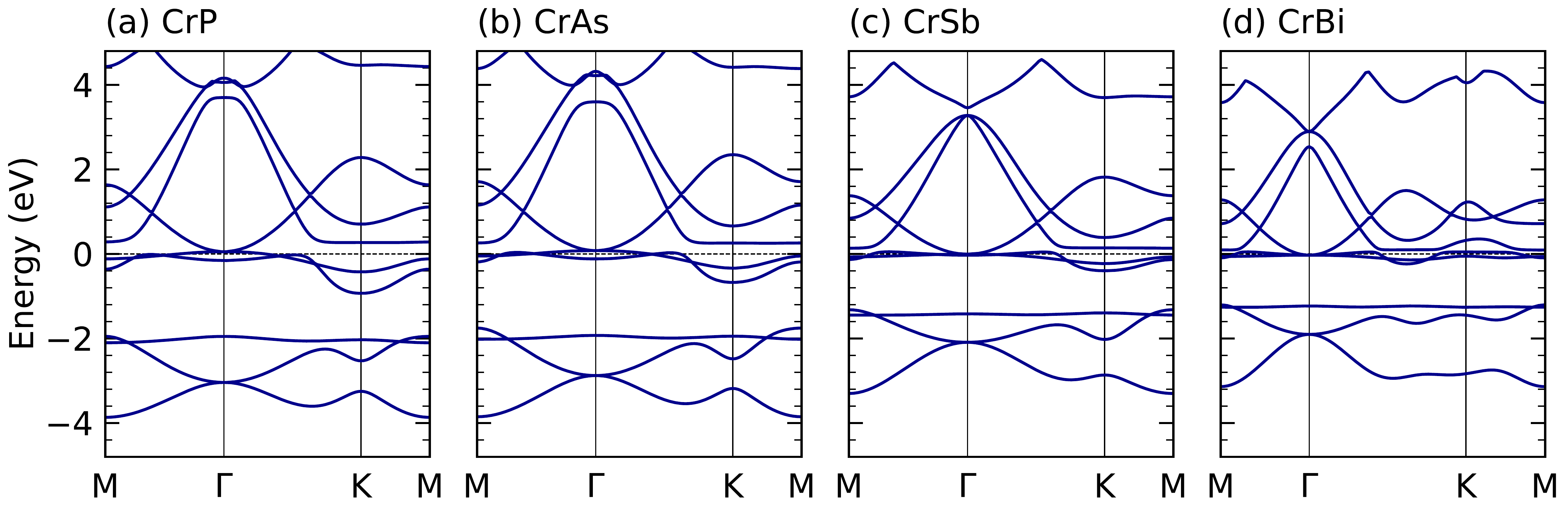}
  \caption{Electronic band structure for spin-unpolarized Cr$X$: 
  (a) CrP, (b) CrAs, (c) CrSb, and (d) CrBi.}
\label{fig:BandNonSpin}
\end{figure*}

In Figs.~\ref{fig:BandNonSpin}(a)--(d), we show the electronic band structure for non-spin polarized Cr$X$, calculated by the {\sc{Quantum ESPRESSO}} package~\cite{QE} with the following DFT details. We set a vacuum layer to 30 \AA{} to avoid the interlayer interactions due to the lattice periodicity. We employ the optimized norm-conserving Vanderbilt (ONCV) pseudopotentials~\cite{hamann2013, schlipf2015} to describe the interaction between electrons and ions. The Perdew-Burke-Ernzerhof (PBE) functional~\cite{pbe1996} under generalized gradient approximation (GGA) is used to describe exchange-correlation energy and potential. We sample the Brillouin zone using a dense $32\times32\times1$ Monkhorst-Pack grid for scf calculation. The plane-wave basis set is used to expand wave functions with cutoff energy of $\sim$820 eV. 

\newpage

\section{Phonon dispersion}

\begin{figure*}[h!]
  \centering \includegraphics[width=0.85\textwidth]{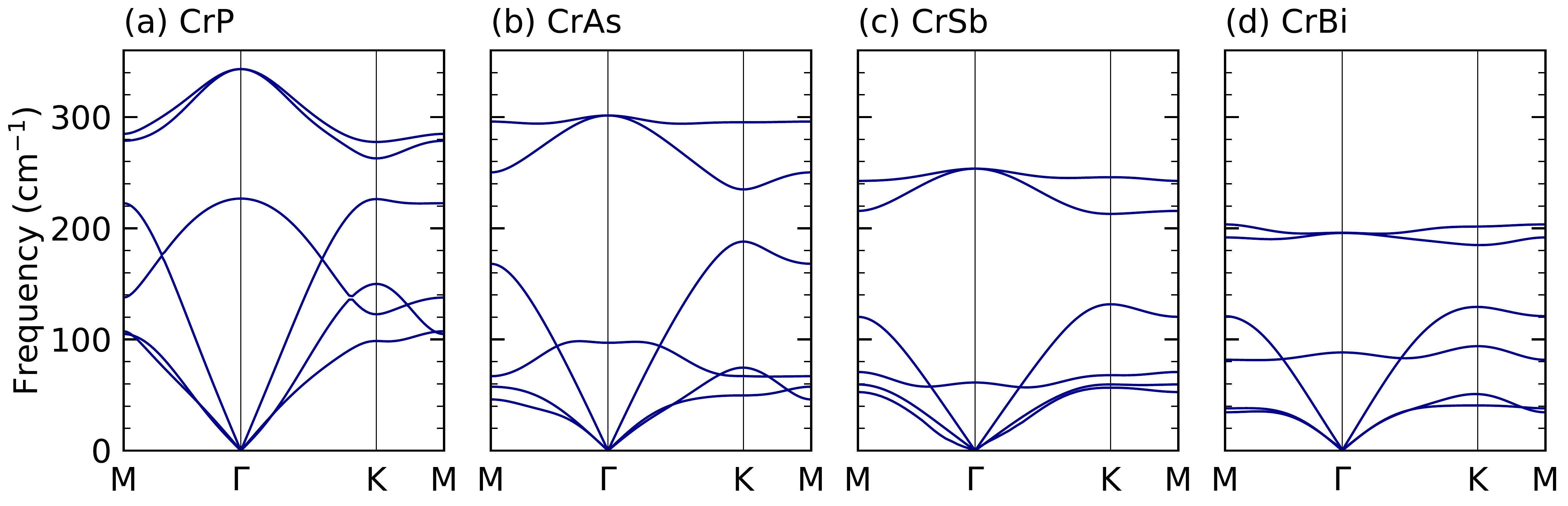}
  \caption{Phonon dispersion relations of spin-polarized Cr$X$:
  (a) CrP, (b) CrAs, (c) CrSb), and (d) CrBi.}
\label{fig:Phonon}
\end{figure*}

In Figs.~\ref{fig:Phonon}(a)--(d), we show the phonon dispersion relations for CrX from the {\texttt{ph.x}} calculations in the {\sc{Quantum ESPRESSO}} package with the dynamical matrix integral of $4\times4\times1$ mesh of \textbf{q}-points. We found no negative and no imaginary frequency of phonons for all species of Cr$X$, thus indicating the dynamical stability of the calculated structures. Unlike their electronic structures, phonon band structures of chromium pnictides resemble that of graphene. There are six phonon modes comes from two atoms in a unit cell with three polarization directions. Heterogeneous atoms in a unit cell splits the in-plane optical modes located at the top branches i.e., longitudinal optic and in-plane tangential optic modes. These two modes are degenerate in graphene at $\Gamma$ point.  The highest optical branch is, interestingly, flat in CrAs, CrSb, and CrBi. As atomic number increases from CrP to CrBi, phonon bandwidth and sound velocity decrease as expected from harmonic oscillator model. This feature will play roles in controlling thermal conductivity of chromium pnictides.

\section{Relaxation time model}
\begin{figure*}[h!]
  \centering \includegraphics[width=0.85\textwidth]{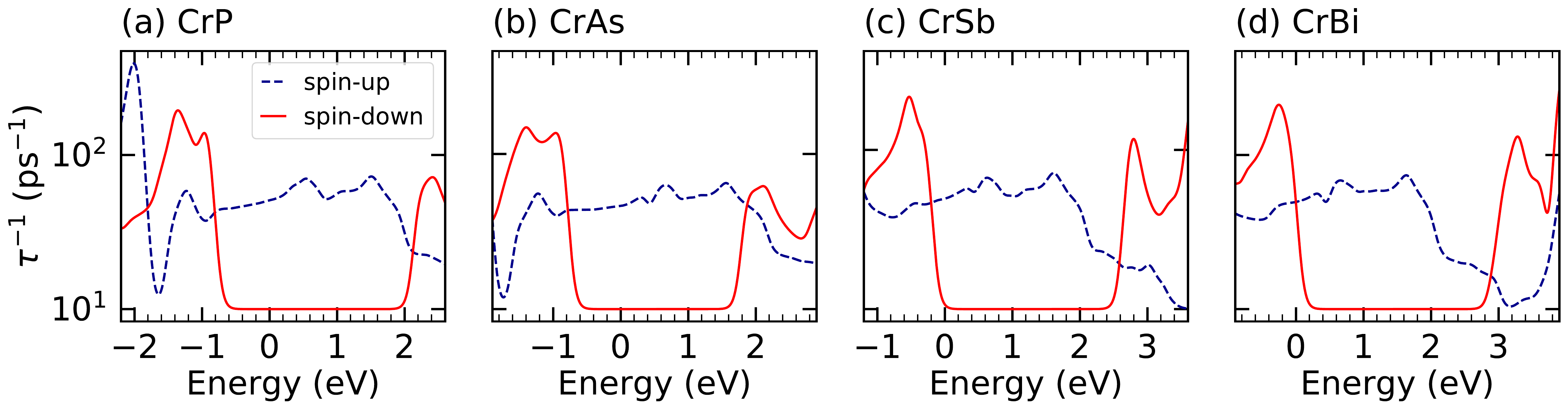}
  \caption{Inverse relaxation time or scattering rate model as
  a function of energy for each Cr$X$:
  (a) CrP, (b) CrAs, (c) CrSb, and (d) CrBi.}
\label{fig:InvTau}
\end{figure*}

We include the electron-phonon scattering rate in the relaxation time, that can be defined as $\tau_{n,\textbf{k}}=\hbar/(2~\mathrm{Im}\sum_{n,\textbf{k}})$, where $\hbar$ is the reduced Planck constant and $\mathrm{Im}\sum_{n,\textbf{k}}$ is the imaginary part of the electronic self-energy, as implemented in the {\sc{EPW}} code \cite{epw}. We compute the electron energy (and the phonon dispersion) on a relatively coarse $16\times16\times1$ ($8\times8\times1$) \textbf{k}-point (\textbf{q}-point) grid. To obtain a finer result, we fit $1/\tau_{n,\textbf{k}}$ from {\sc{EPW}} with an energy-dependent relaxation time: $1/\tau(\varepsilon)=\mathcal{C}\cdot {\mathrm{DOS}}(\varepsilon)+ 1/\tau_0$, where $\mathcal{C}$ is a DOS-dependent fitting parameter and $\tau_0$ is the relaxation time constant that is chosen to overcome the absence of $\mathrm{DOS}$ in the band gap area in the spin-down band.  For all Cr$X$ monolayers, the fitting parameter is set to $\mathcal{C}=10^{12}$ eV/s and $\tau_0=10^{-13}$ s. The inverse relaxation time as a function of energy for Cr$X$ are shown in Figs.~\ref{fig:InvTau}(a)--(d).

\section{Thermoelectric transport coefficients}
\begin{figure*}[t]
  \centering \includegraphics[width=0.85\textwidth]{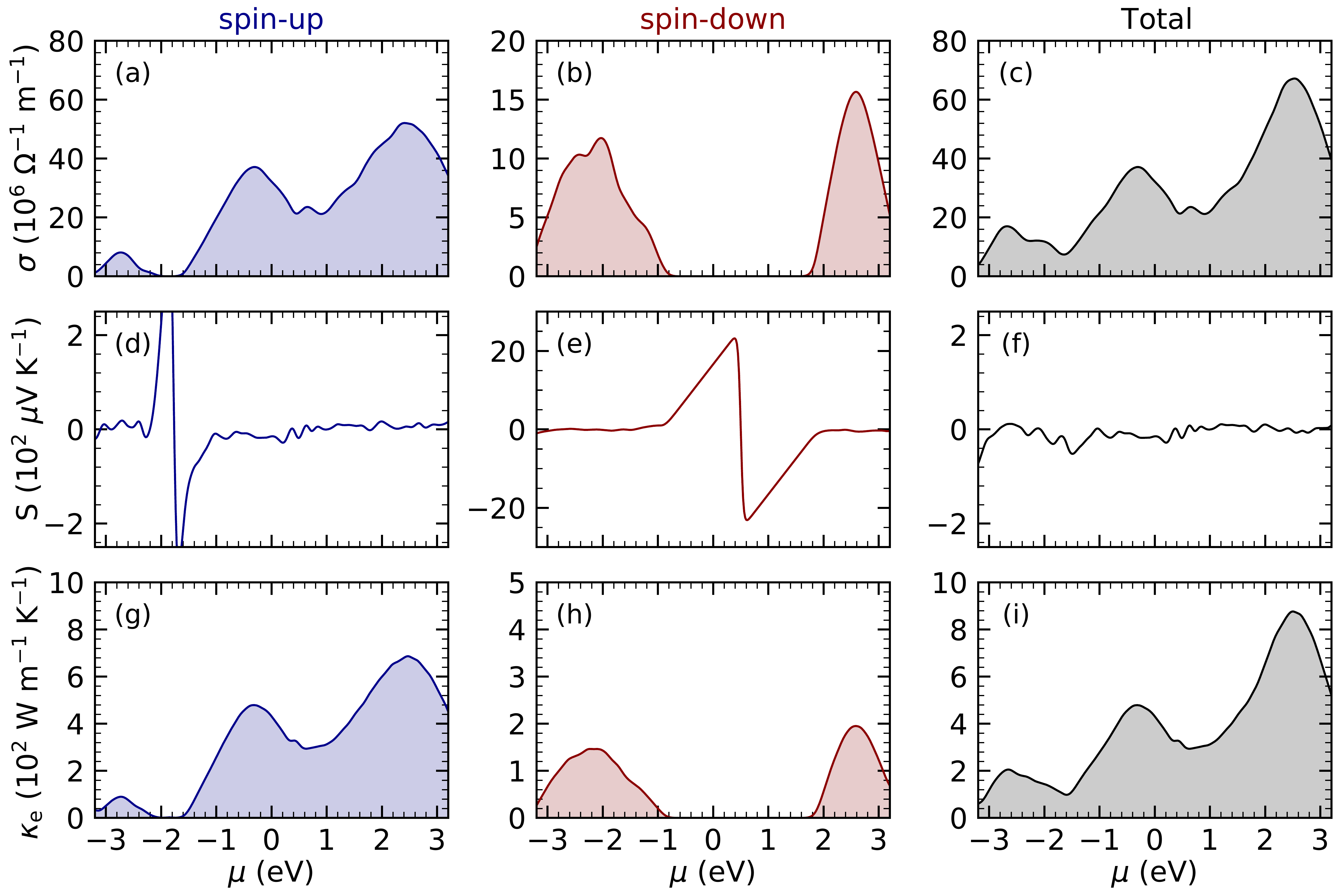}
  \caption{Thermoelectric transport coefficients of CrAs.}
\label{fig:TE-CrAs}
\end{figure*}

\begin{figure*}[t]
  \centering \includegraphics[width=0.85\textwidth]{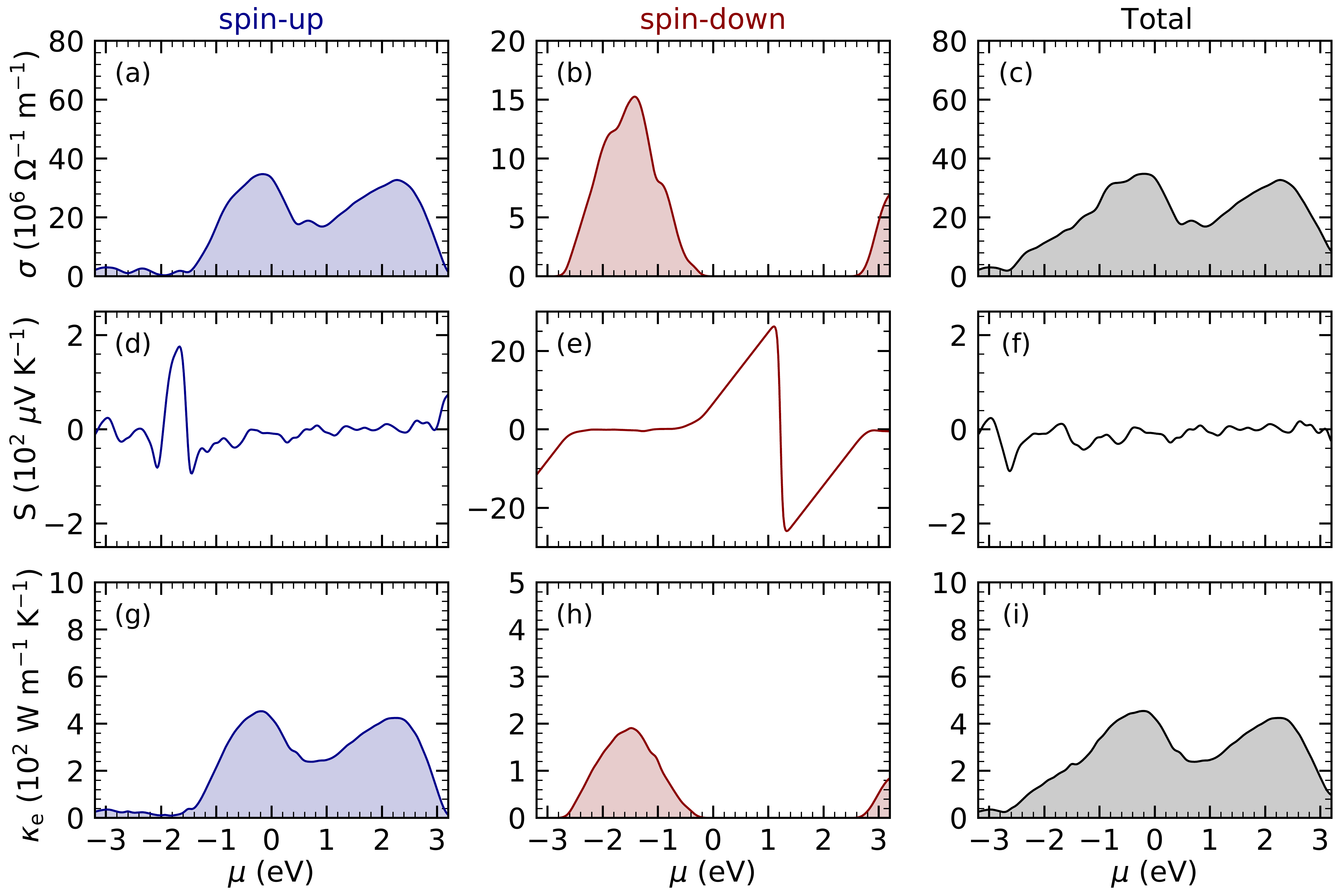}
  \caption{Thermoelectric transport coefficients of CrSb.}
\label{fig:TE-CrSb}
\end{figure*}

\begin{figure*}[t]
  \centering \includegraphics[width=0.85\textwidth]{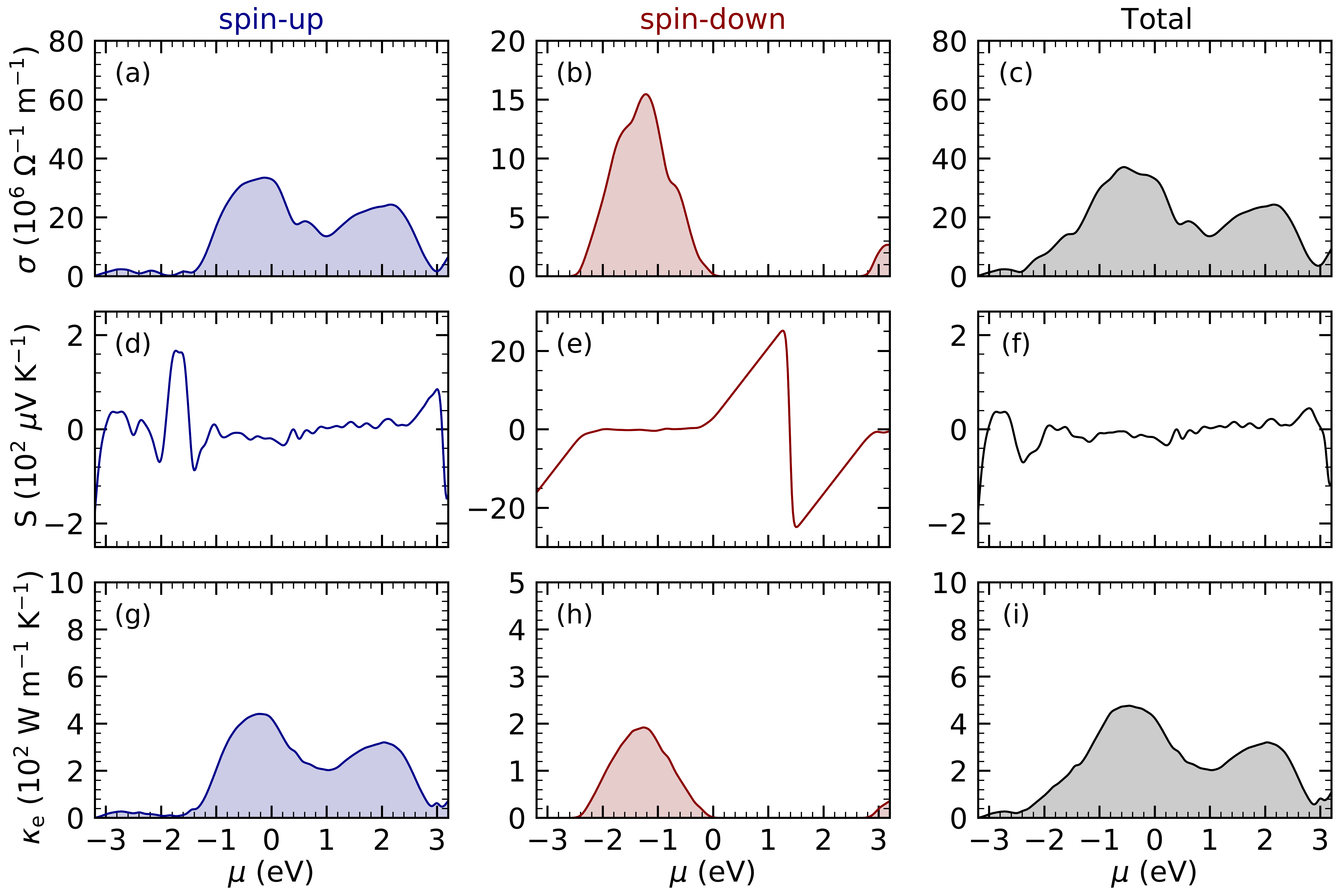}
  \caption{Thermoelectric transport coefficients of CrBi.}
\label{fig:TE-CrBi}
\end{figure*}

In Fig.~\ref{fig:TE-CrAs}--\ref{fig:TE-CrBi}, we plot the thermoelectric transport coefficients as a function of chemical potential of CrAs, CrSb, and CrBi at $T=500~\rm K$.  The thermoelectric properties are calculated using {\sc{BoltzTraP2}} code~\cite{boltztrap2} within Boltzmann transport equation under the energy-dependent relaxation time approximation. 

\begin{thebibliography}{43}%
\makeatletter
\providecommand \@ifxundefined [1]{%
 \@ifx{#1\undefined}
}%
\providecommand \@ifnum [1]{%
 \ifnum #1\expandafter \@firstoftwo
 \else \expandafter \@secondoftwo
 \fi
}%
\providecommand \@ifx [1]{%
 \ifx #1\expandafter \@firstoftwo
 \else \expandafter \@secondoftwo
 \fi
}%
\providecommand \natexlab [1]{#1}%
\providecommand \enquote  [1]{``#1''}%
\providecommand \bibnamefont  [1]{#1}%
\providecommand \bibfnamefont [1]{#1}%
\providecommand \citenamefont [1]{#1}%
\providecommand \href@noop [0]{\@secondoftwo}%
\providecommand \href [0]{\begingroup \@sanitize@url \@href}%
\providecommand \@href[1]{\@@startlink{#1}\@@href}%
\providecommand \@@href[1]{\endgroup#1\@@endlink}%
\providecommand \@sanitize@url [0]{\catcode `\\12\catcode `\$12\catcode
  `\&12\catcode `\#12\catcode `\^12\catcode `\_12\catcode `\%12\relax}%
\providecommand \@@startlink[1]{}%
\providecommand \@@endlink[0]{}%
\providecommand \url  [0]{\begingroup\@sanitize@url \@url }%
\providecommand \@url [1]{\endgroup\@href {#1}{\urlprefix }}%
\providecommand \urlprefix  [0]{URL }%
\providecommand \Eprint [0]{\href }%
\providecommand \doibase [0]{https://doi.org/}%
\providecommand \selectlanguage [0]{\@gobble}%
\providecommand \bibinfo  [0]{\@secondoftwo}%
\providecommand \bibfield  [0]{\@secondoftwo}%
\providecommand \translation [1]{[#1]}%
\providecommand \BibitemOpen [0]{}%
\providecommand \bibitemStop [0]{}%
\providecommand \bibitemNoStop [0]{.\EOS\space}%
\providecommand \EOS [0]{\spacefactor3000\relax}%
\providecommand \BibitemShut  [1]{\csname bibitem#1\endcsname}%
\let\auto@bib@innerbib\@empty
\bibitem [{\citenamefont {Goldsmid}(2010)}]{goldsmid2010}%
  \BibitemOpen
  \bibfield  {author} {\bibinfo {author} {\bibfnamefont {H.~J.}\ \bibnamefont
  {Goldsmid}},\ }\href@noop {} {\emph {\bibinfo {title} {Introduction to
  Thermoelectricity}}}\ (\bibinfo  {publisher} {Springer-Verlag},\ \bibinfo
  {address} {Berlin Heidelberg},\ \bibinfo {year} {2010})\BibitemShut {NoStop}%
\bibitem [{\citenamefont {Liu}\ \emph {et~al.}(2016)\citenamefont {Liu},
  \citenamefont {Kim}, \citenamefont {Jie},\ and\ \citenamefont
  {Ren}}]{liu2016}%
  \BibitemOpen
  \bibfield  {author} {\bibinfo {author} {\bibfnamefont {W.}~\bibnamefont
  {Liu}}, \bibinfo {author} {\bibfnamefont {H.~S.}\ \bibnamefont {Kim}},
  \bibinfo {author} {\bibfnamefont {Q.}~\bibnamefont {Jie}},\ and\ \bibinfo
  {author} {\bibfnamefont {Z.}~\bibnamefont {Ren}},\ }\href@noop {} {\bibfield
  {journal} {\bibinfo  {journal} {Scripta Mater.}\ }\textbf {\bibinfo {volume}
  {111}},\ \bibinfo {pages} {3} (\bibinfo {year} {2016})}\BibitemShut {NoStop}%
\bibitem [{\citenamefont {Minnich}\ \emph {et~al.}(2009)\citenamefont
  {Minnich}, \citenamefont {Dresselhaus}, \citenamefont {Ren},\ and\
  \citenamefont {Chen}}]{minnich2009}%
  \BibitemOpen
  \bibfield  {author} {\bibinfo {author} {\bibfnamefont {A.~J.}\ \bibnamefont
  {Minnich}}, \bibinfo {author} {\bibfnamefont {M.~S.}\ \bibnamefont
  {Dresselhaus}}, \bibinfo {author} {\bibfnamefont {Z.~F.}\ \bibnamefont
  {Ren}},\ and\ \bibinfo {author} {\bibfnamefont {G.}~\bibnamefont {Chen}},\
  }\href@noop {} {\bibfield  {journal} {\bibinfo  {journal} {Energy Environ.
  Sci.}\ }\textbf {\bibinfo {volume} {2}},\ \bibinfo {pages} {466} (\bibinfo
  {year} {2009})}\BibitemShut {NoStop}%
\bibitem [{\citenamefont {Zebarjadi}\ \emph {et~al.}(2012)\citenamefont
  {Zebarjadi}, \citenamefont {Esfarjani}, \citenamefont {Dresselhaus},
  \citenamefont {Ren},\ and\ \citenamefont {Chen}}]{mona2012}%
  \BibitemOpen
  \bibfield  {author} {\bibinfo {author} {\bibfnamefont {M.}~\bibnamefont
  {Zebarjadi}}, \bibinfo {author} {\bibfnamefont {K.}~\bibnamefont
  {Esfarjani}}, \bibinfo {author} {\bibfnamefont {M.~S.}\ \bibnamefont
  {Dresselhaus}}, \bibinfo {author} {\bibfnamefont {Z.~F.}\ \bibnamefont
  {Ren}},\ and\ \bibinfo {author} {\bibfnamefont {G.}~\bibnamefont {Chen}},\
  }\href@noop {} {\bibfield  {journal} {\bibinfo  {journal} {Energy Environ.
  Sci.}\ }\textbf {\bibinfo {volume} {5}},\ \bibinfo {pages} {5147} (\bibinfo
  {year} {2012})}\BibitemShut {NoStop}%
\bibitem [{\citenamefont {Heremans}\ \emph {et~al.}(2013)\citenamefont
  {Heremans}, \citenamefont {Dresselhaus}, \citenamefont {Bell},\ and\
  \citenamefont {Morelli}}]{heremans2013}%
  \BibitemOpen
  \bibfield  {author} {\bibinfo {author} {\bibfnamefont {J.~P.}\ \bibnamefont
  {Heremans}}, \bibinfo {author} {\bibfnamefont {M.~S.}\ \bibnamefont
  {Dresselhaus}}, \bibinfo {author} {\bibfnamefont {L.~E.}\ \bibnamefont
  {Bell}},\ and\ \bibinfo {author} {\bibfnamefont {D.~T.}\ \bibnamefont
  {Morelli}},\ }\href@noop {} {\bibfield  {journal} {\bibinfo  {journal} {Nat.
  Nanotechnol.}\ }\textbf {\bibinfo {volume} {8}},\ \bibinfo {pages} {471}
  (\bibinfo {year} {2013})}\BibitemShut {NoStop}%
\bibitem [{\citenamefont {Franz}\ and\ \citenamefont
  {Wiedemann}(1853)}]{wflaw}%
  \BibitemOpen
  \bibfield  {author} {\bibinfo {author} {\bibfnamefont {R.}~\bibnamefont
  {Franz}}\ and\ \bibinfo {author} {\bibfnamefont {G.}~\bibnamefont
  {Wiedemann}},\ }\href@noop {} {\bibfield  {journal} {\bibinfo  {journal}
  {Ann. Phys.}\ }\textbf {\bibinfo {volume} {165}},\ \bibinfo {pages} {497}
  (\bibinfo {year} {1853})}\BibitemShut {NoStop}%
\bibitem [{\citenamefont {Shakouri}(2011)}]{shakouri2011}%
  \BibitemOpen
  \bibfield  {author} {\bibinfo {author} {\bibfnamefont {A.}~\bibnamefont
  {Shakouri}},\ }\href@noop {} {\bibfield  {journal} {\bibinfo  {journal}
  {Annu. Rev. Mater. Res.}\ }\textbf {\bibinfo {volume} {41}},\ \bibinfo
  {pages} {399} (\bibinfo {year} {2011})}\BibitemShut {NoStop}%
\bibitem [{\citenamefont {Dresselhaus}\ \emph {et~al.}(2007)\citenamefont
  {Dresselhaus}, \citenamefont {Chen}, \citenamefont {Tang}, \citenamefont
  {Yang}, \citenamefont {Lee}, \citenamefont {Wang}, \citenamefont {Ren},
  \citenamefont {Fleurial},\ and\ \citenamefont {Gogna}}]{dresselhaus2007}%
  \BibitemOpen
  \bibfield  {author} {\bibinfo {author} {\bibfnamefont {M.}~\bibnamefont
  {Dresselhaus}}, \bibinfo {author} {\bibfnamefont {G.}~\bibnamefont {Chen}},
  \bibinfo {author} {\bibfnamefont {M.}~\bibnamefont {Tang}}, \bibinfo {author}
  {\bibfnamefont {R.}~\bibnamefont {Yang}}, \bibinfo {author} {\bibfnamefont
  {H.}~\bibnamefont {Lee}}, \bibinfo {author} {\bibfnamefont {D.}~\bibnamefont
  {Wang}}, \bibinfo {author} {\bibfnamefont {Z.}~\bibnamefont {Ren}}, \bibinfo
  {author} {\bibfnamefont {J.-P.}\ \bibnamefont {Fleurial}},\ and\ \bibinfo
  {author} {\bibfnamefont {P.}~\bibnamefont {Gogna}},\ }\href@noop {}
  {\bibfield  {journal} {\bibinfo  {journal} {Adv. Mater.}\ }\textbf {\bibinfo
  {volume} {19}},\ \bibinfo {pages} {1043} (\bibinfo {year}
  {2007})}\BibitemShut {NoStop}%
\bibitem [{\citenamefont {Markov}\ \emph {et~al.}(2018)\citenamefont {Markov},
  \citenamefont {Hu}, \citenamefont {Liu}, \citenamefont {Liu}, \citenamefont
  {Poon}, \citenamefont {Esfarjani},\ and\ \citenamefont
  {Zebarjadi}}]{markov2018}%
  \BibitemOpen
  \bibfield  {author} {\bibinfo {author} {\bibfnamefont {M.}~\bibnamefont
  {Markov}}, \bibinfo {author} {\bibfnamefont {X.}~\bibnamefont {Hu}}, \bibinfo
  {author} {\bibfnamefont {H.-C.}\ \bibnamefont {Liu}}, \bibinfo {author}
  {\bibfnamefont {N.}~\bibnamefont {Liu}}, \bibinfo {author} {\bibfnamefont
  {S.~J.}\ \bibnamefont {Poon}}, \bibinfo {author} {\bibfnamefont
  {K.}~\bibnamefont {Esfarjani}},\ and\ \bibinfo {author} {\bibfnamefont
  {M.}~\bibnamefont {Zebarjadi}},\ }\href@noop {} {\bibfield  {journal}
  {\bibinfo  {journal} {Sci. Rep.}\ }\textbf {\bibinfo {volume} {8}},\ \bibinfo
  {pages} {1} (\bibinfo {year} {2018})}\BibitemShut {NoStop}%
\bibitem [{\citenamefont {Markov}\ and\ \citenamefont
  {Zebarjadi}(2019)}]{markov2019}%
  \BibitemOpen
  \bibfield  {author} {\bibinfo {author} {\bibfnamefont {M.}~\bibnamefont
  {Markov}}\ and\ \bibinfo {author} {\bibfnamefont {M.}~\bibnamefont
  {Zebarjadi}},\ }\href@noop {} {\bibfield  {journal} {\bibinfo  {journal}
  {Nanoscale Microscale Thermophys. Eng.}\ }\textbf {\bibinfo {volume} {23}},\
  \bibinfo {pages} {117} (\bibinfo {year} {2019})}\BibitemShut {NoStop}%
\bibitem [{\citenamefont {Kanahashi}\ \emph {et~al.}(2020)\citenamefont
  {Kanahashi}, \citenamefont {Pu},\ and\ \citenamefont
  {Takenobu}}]{kanahasi2020}%
  \BibitemOpen
  \bibfield  {author} {\bibinfo {author} {\bibfnamefont {K.}~\bibnamefont
  {Kanahashi}}, \bibinfo {author} {\bibfnamefont {J.}~\bibnamefont {Pu}},\ and\
  \bibinfo {author} {\bibfnamefont {T.}~\bibnamefont {Takenobu}},\ }\href@noop
  {} {\bibfield  {journal} {\bibinfo  {journal} {Adv. Energy Mater.}\ }\textbf
  {\bibinfo {volume} {10}},\ \bibinfo {pages} {1902842} (\bibinfo {year}
  {2020})}\BibitemShut {NoStop}%
\bibitem [{\citenamefont {Li}\ \emph {et~al.}(2020)\citenamefont {Li},
  \citenamefont {Gong}, \citenamefont {Chen}, \citenamefont {Lin},
  \citenamefont {Khan}, \citenamefont {Zhang}, \citenamefont {Li},
  \citenamefont {Zhang},\ and\ \citenamefont {Xie}}]{li2020}%
  \BibitemOpen
  \bibfield  {author} {\bibinfo {author} {\bibfnamefont {D.}~\bibnamefont
  {Li}}, \bibinfo {author} {\bibfnamefont {Y.}~\bibnamefont {Gong}}, \bibinfo
  {author} {\bibfnamefont {Y.}~\bibnamefont {Chen}}, \bibinfo {author}
  {\bibfnamefont {J.}~\bibnamefont {Lin}}, \bibinfo {author} {\bibfnamefont
  {Q.}~\bibnamefont {Khan}}, \bibinfo {author} {\bibfnamefont {Y.}~\bibnamefont
  {Zhang}}, \bibinfo {author} {\bibfnamefont {Y.}~\bibnamefont {Li}}, \bibinfo
  {author} {\bibfnamefont {H.}~\bibnamefont {Zhang}},\ and\ \bibinfo {author}
  {\bibfnamefont {H.}~\bibnamefont {Xie}},\ }\href@noop {} {\bibfield
  {journal} {\bibinfo  {journal} {Nano\text{-}Micro Lett.}\ }\textbf {\bibinfo
  {volume} {12}},\ \bibinfo {pages} {1} (\bibinfo {year} {2020})}\BibitemShut
  {NoStop}%
\bibitem [{\citenamefont {Fei}\ \emph {et~al.}(2014)\citenamefont {Fei},
  \citenamefont {Faghaninia}, \citenamefont {Soklaski}, \citenamefont {Yan},
  \citenamefont {Lo},\ and\ \citenamefont {Yang}}]{fei2014}%
  \BibitemOpen
  \bibfield  {author} {\bibinfo {author} {\bibfnamefont {R.}~\bibnamefont
  {Fei}}, \bibinfo {author} {\bibfnamefont {A.}~\bibnamefont {Faghaninia}},
  \bibinfo {author} {\bibfnamefont {R.}~\bibnamefont {Soklaski}}, \bibinfo
  {author} {\bibfnamefont {J.-A.}\ \bibnamefont {Yan}}, \bibinfo {author}
  {\bibfnamefont {C.}~\bibnamefont {Lo}},\ and\ \bibinfo {author}
  {\bibfnamefont {L.}~\bibnamefont {Yang}},\ }\href@noop {} {\bibfield
  {journal} {\bibinfo  {journal} {Nano Lett.}\ }\textbf {\bibinfo {volume}
  {14}},\ \bibinfo {pages} {6393} (\bibinfo {year} {2014})}\BibitemShut
  {NoStop}%
\bibitem [{\citenamefont {Liao}\ \emph {et~al.}(2015)\citenamefont {Liao},
  \citenamefont {Zhou}, \citenamefont {Qiu}, \citenamefont {Dresselhaus},\ and\
  \citenamefont {Chen}}]{bolin2015}%
  \BibitemOpen
  \bibfield  {author} {\bibinfo {author} {\bibfnamefont {B.}~\bibnamefont
  {Liao}}, \bibinfo {author} {\bibfnamefont {J.}~\bibnamefont {Zhou}}, \bibinfo
  {author} {\bibfnamefont {B.}~\bibnamefont {Qiu}}, \bibinfo {author}
  {\bibfnamefont {M.~S.}\ \bibnamefont {Dresselhaus}},\ and\ \bibinfo {author}
  {\bibfnamefont {G.}~\bibnamefont {Chen}},\ }\href@noop {} {\bibfield
  {journal} {\bibinfo  {journal} {Phys. Rev. B}\ }\textbf {\bibinfo {volume}
  {91}},\ \bibinfo {pages} {235419} (\bibinfo {year} {2015})}\BibitemShut
  {NoStop}%
\bibitem [{\citenamefont {Saito}\ \emph {et~al.}(2016)\citenamefont {Saito},
  \citenamefont {Iizuka}, \citenamefont {Koretsune}, \citenamefont {Arita},
  \citenamefont {Shimizu},\ and\ \citenamefont {Iwasa}}]{iwasa2016}%
  \BibitemOpen
  \bibfield  {author} {\bibinfo {author} {\bibfnamefont {Y.}~\bibnamefont
  {Saito}}, \bibinfo {author} {\bibfnamefont {T.}~\bibnamefont {Iizuka}},
  \bibinfo {author} {\bibfnamefont {T.}~\bibnamefont {Koretsune}}, \bibinfo
  {author} {\bibfnamefont {R.}~\bibnamefont {Arita}}, \bibinfo {author}
  {\bibfnamefont {S.}~\bibnamefont {Shimizu}},\ and\ \bibinfo {author}
  {\bibfnamefont {Y.}~\bibnamefont {Iwasa}},\ }\href@noop {} {\bibfield
  {journal} {\bibinfo  {journal} {Nano Lett.}\ }\textbf {\bibinfo {volume}
  {16}},\ \bibinfo {pages} {4819} (\bibinfo {year} {2016})}\BibitemShut
  {NoStop}%
\bibitem [{\citenamefont {Jin}\ \emph {et~al.}(2015)\citenamefont {Jin},
  \citenamefont {Liao}, \citenamefont {Fang}, \citenamefont {Liu},
  \citenamefont {Liu}, \citenamefont {Ding}, \citenamefont {Luo},\ and\
  \citenamefont {Yang}}]{jin2015}%
  \BibitemOpen
  \bibfield  {author} {\bibinfo {author} {\bibfnamefont {Z.}~\bibnamefont
  {Jin}}, \bibinfo {author} {\bibfnamefont {Q.}~\bibnamefont {Liao}}, \bibinfo
  {author} {\bibfnamefont {H.}~\bibnamefont {Fang}}, \bibinfo {author}
  {\bibfnamefont {Z.}~\bibnamefont {Liu}}, \bibinfo {author} {\bibfnamefont
  {W.}~\bibnamefont {Liu}}, \bibinfo {author} {\bibfnamefont {Z.}~\bibnamefont
  {Ding}}, \bibinfo {author} {\bibfnamefont {T.}~\bibnamefont {Luo}},\ and\
  \bibinfo {author} {\bibfnamefont {N.}~\bibnamefont {Yang}},\ }\href@noop {}
  {\bibfield  {journal} {\bibinfo  {journal} {Sci. Rep.}\ }\textbf {\bibinfo
  {volume} {5}},\ \bibinfo {pages} {1} (\bibinfo {year} {2015})}\BibitemShut
  {NoStop}%
\bibitem [{\citenamefont {Yoshida}\ \emph {et~al.}(2016)\citenamefont
  {Yoshida}, \citenamefont {Iizuka}, \citenamefont {Saito}, \citenamefont
  {Onga}, \citenamefont {Suzuki}, \citenamefont {Zhang}, \citenamefont
  {Iwasa},\ and\ \citenamefont {Shimizu}}]{yoshida2016}%
  \BibitemOpen
  \bibfield  {author} {\bibinfo {author} {\bibfnamefont {M.}~\bibnamefont
  {Yoshida}}, \bibinfo {author} {\bibfnamefont {T.}~\bibnamefont {Iizuka}},
  \bibinfo {author} {\bibfnamefont {Y.}~\bibnamefont {Saito}}, \bibinfo
  {author} {\bibfnamefont {M.}~\bibnamefont {Onga}}, \bibinfo {author}
  {\bibfnamefont {R.}~\bibnamefont {Suzuki}}, \bibinfo {author} {\bibfnamefont
  {Y.}~\bibnamefont {Zhang}}, \bibinfo {author} {\bibfnamefont
  {Y.}~\bibnamefont {Iwasa}},\ and\ \bibinfo {author} {\bibfnamefont
  {S.}~\bibnamefont {Shimizu}},\ }\href@noop {} {\bibfield  {journal} {\bibinfo
   {journal} {Nano Lett.}\ }\textbf {\bibinfo {volume} {16}},\ \bibinfo {pages}
  {2061} (\bibinfo {year} {2016})}\BibitemShut {NoStop}%
\bibitem [{\citenamefont {Wickramaratne}\ \emph {et~al.}(2015)\citenamefont
  {Wickramaratne}, \citenamefont {Zahid},\ and\ \citenamefont
  {Lake}}]{wickramaratne2015}%
  \BibitemOpen
  \bibfield  {author} {\bibinfo {author} {\bibfnamefont {D.}~\bibnamefont
  {Wickramaratne}}, \bibinfo {author} {\bibfnamefont {F.}~\bibnamefont
  {Zahid}},\ and\ \bibinfo {author} {\bibfnamefont {R.~K.}\ \bibnamefont
  {Lake}},\ }\href@noop {} {\bibfield  {journal} {\bibinfo  {journal} {J. Appl.
  Phys.}\ }\textbf {\bibinfo {volume} {118}},\ \bibinfo {pages} {075101}
  (\bibinfo {year} {2015})}\BibitemShut {NoStop}%
\bibitem [{\citenamefont {Zeng}\ \emph {et~al.}(2018)\citenamefont {Zeng},
  \citenamefont {He}, \citenamefont {Liang}, \citenamefont {Liu}, \citenamefont
  {Sun}, \citenamefont {Pan}, \citenamefont {Wang}, \citenamefont {Cao},
  \citenamefont {Liu}, \citenamefont {Wang}, \citenamefont {Zhang},
  \citenamefont {Yan}, \citenamefont {Su}, \citenamefont {Wang}, \citenamefont
  {Watanabe}, \citenamefont {Taniguchi}, \citenamefont {Singh}, \citenamefont
  {Zhang},\ and\ \citenamefont {Miao}}]{zeng2018}%
  \BibitemOpen
  \bibfield  {author} {\bibinfo {author} {\bibfnamefont {J.}~\bibnamefont
  {Zeng}}, \bibinfo {author} {\bibfnamefont {X.}~\bibnamefont {He}}, \bibinfo
  {author} {\bibfnamefont {S.-J.}\ \bibnamefont {Liang}}, \bibinfo {author}
  {\bibfnamefont {E.}~\bibnamefont {Liu}}, \bibinfo {author} {\bibfnamefont
  {Y.}~\bibnamefont {Sun}}, \bibinfo {author} {\bibfnamefont {C.}~\bibnamefont
  {Pan}}, \bibinfo {author} {\bibfnamefont {Y.}~\bibnamefont {Wang}}, \bibinfo
  {author} {\bibfnamefont {T.}~\bibnamefont {Cao}}, \bibinfo {author}
  {\bibfnamefont {X.}~\bibnamefont {Liu}}, \bibinfo {author} {\bibfnamefont
  {C.}~\bibnamefont {Wang}}, \bibinfo {author} {\bibfnamefont {L.}~\bibnamefont
  {Zhang}}, \bibinfo {author} {\bibfnamefont {S.}~\bibnamefont {Yan}}, \bibinfo
  {author} {\bibfnamefont {G.}~\bibnamefont {Su}}, \bibinfo {author}
  {\bibfnamefont {Z.}~\bibnamefont {Wang}}, \bibinfo {author} {\bibfnamefont
  {K.}~\bibnamefont {Watanabe}}, \bibinfo {author} {\bibfnamefont
  {T.}~\bibnamefont {Taniguchi}}, \bibinfo {author} {\bibfnamefont {D.~J.}\
  \bibnamefont {Singh}}, \bibinfo {author} {\bibfnamefont {L.}~\bibnamefont
  {Zhang}},\ and\ \bibinfo {author} {\bibfnamefont {F.}~\bibnamefont {Miao}},\
  }\href@noop {} {\bibfield  {journal} {\bibinfo  {journal} {Nano Lett.}\
  }\textbf {\bibinfo {volume} {18}},\ \bibinfo {pages} {7538} (\bibinfo {year}
  {2018})}\BibitemShut {NoStop}%
\bibitem [{\citenamefont {Hicks}\ and\ \citenamefont
  {Dresselhaus}(1993)}]{hicks93}%
  \BibitemOpen
  \bibfield  {author} {\bibinfo {author} {\bibfnamefont {L.~D.}\ \bibnamefont
  {Hicks}}\ and\ \bibinfo {author} {\bibfnamefont {M.~S.}\ \bibnamefont
  {Dresselhaus}},\ }\href@noop {} {\bibfield  {journal} {\bibinfo  {journal}
  {Phys. Rev. B}\ }\textbf {\bibinfo {volume} {47}},\ \bibinfo {pages} {12727}
  (\bibinfo {year} {1993})}\BibitemShut {NoStop}%
\bibitem [{\citenamefont {Hicks}\ \emph {et~al.}(1996)\citenamefont {Hicks},
  \citenamefont {Harman}, \citenamefont {Sun},\ and\ \citenamefont
  {Dresselhaus}}]{hicks96}%
  \BibitemOpen
  \bibfield  {author} {\bibinfo {author} {\bibfnamefont {L.~D.}\ \bibnamefont
  {Hicks}}, \bibinfo {author} {\bibfnamefont {T.~C.}\ \bibnamefont {Harman}},
  \bibinfo {author} {\bibfnamefont {X.}~\bibnamefont {Sun}},\ and\ \bibinfo
  {author} {\bibfnamefont {M.~S.}\ \bibnamefont {Dresselhaus}},\ }\href@noop {}
  {\bibfield  {journal} {\bibinfo  {journal} {Phys. Rev. B}\ }\textbf {\bibinfo
  {volume} {53}},\ \bibinfo {pages} {R10493} (\bibinfo {year}
  {1996})}\BibitemShut {NoStop}%
\bibitem [{\citenamefont {Hung}\ \emph {et~al.}(2016)\citenamefont {Hung},
  \citenamefont {Hasdeo}, \citenamefont {Nugraha}, \citenamefont
  {Dresselhaus},\ and\ \citenamefont {Saito}}]{hung2016-prl}%
  \BibitemOpen
  \bibfield  {author} {\bibinfo {author} {\bibfnamefont {N.~T.}\ \bibnamefont
  {Hung}}, \bibinfo {author} {\bibfnamefont {E.~H.}\ \bibnamefont {Hasdeo}},
  \bibinfo {author} {\bibfnamefont {A.~R.~T.}\ \bibnamefont {Nugraha}},
  \bibinfo {author} {\bibfnamefont {M.~S.}\ \bibnamefont {Dresselhaus}},\ and\
  \bibinfo {author} {\bibfnamefont {R.}~\bibnamefont {Saito}},\ }\href@noop {}
  {\bibfield  {journal} {\bibinfo  {journal} {Phys. Rev. Lett.}\ }\textbf
  {\bibinfo {volume} {117}},\ \bibinfo {pages} {036602} (\bibinfo {year}
  {2016})}\BibitemShut {NoStop}%
\bibitem [{\citenamefont {Mermin}\ and\ \citenamefont
  {Wagner}(1966)}]{merminwagner}%
  \BibitemOpen
  \bibfield  {author} {\bibinfo {author} {\bibfnamefont {N.~D.}\ \bibnamefont
  {Mermin}}\ and\ \bibinfo {author} {\bibfnamefont {H.}~\bibnamefont
  {Wagner}},\ }\href@noop {} {\bibfield  {journal} {\bibinfo  {journal} {Phys.
  Rev. Lett.}\ }\textbf {\bibinfo {volume} {17}},\ \bibinfo {pages} {1133}
  (\bibinfo {year} {1966})}\BibitemShut {NoStop}%
\bibitem [{\citenamefont {Gong}\ \emph {et~al.}(2017)\citenamefont {Gong},
  \citenamefont {Li}, \citenamefont {Li}, \citenamefont {Ji}, \citenamefont
  {Stern}, \citenamefont {Xia}, \citenamefont {Cao}, \citenamefont {Bao},
  \citenamefont {Wang}, \citenamefont {Wang} \emph {et~al.}}]{gong2017}%
  \BibitemOpen
  \bibfield  {author} {\bibinfo {author} {\bibfnamefont {C.}~\bibnamefont
  {Gong}}, \bibinfo {author} {\bibfnamefont {L.}~\bibnamefont {Li}}, \bibinfo
  {author} {\bibfnamefont {Z.}~\bibnamefont {Li}}, \bibinfo {author}
  {\bibfnamefont {H.}~\bibnamefont {Ji}}, \bibinfo {author} {\bibfnamefont
  {A.}~\bibnamefont {Stern}}, \bibinfo {author} {\bibfnamefont
  {Y.}~\bibnamefont {Xia}}, \bibinfo {author} {\bibfnamefont {T.}~\bibnamefont
  {Cao}}, \bibinfo {author} {\bibfnamefont {W.}~\bibnamefont {Bao}}, \bibinfo
  {author} {\bibfnamefont {C.}~\bibnamefont {Wang}}, \bibinfo {author}
  {\bibfnamefont {Y.}~\bibnamefont {Wang}}, \emph {et~al.},\ }\href@noop {}
  {\bibfield  {journal} {\bibinfo  {journal} {Nature}\ }\textbf {\bibinfo
  {volume} {546}},\ \bibinfo {pages} {265} (\bibinfo {year}
  {2017})}\BibitemShut {NoStop}%
\bibitem [{\citenamefont {Huang}\ \emph {et~al.}(2017)\citenamefont {Huang},
  \citenamefont {Clark}, \citenamefont {Navarro-Moratalla}, \citenamefont
  {Klein}, \citenamefont {Cheng}, \citenamefont {Seyler}, \citenamefont
  {Zhong}, \citenamefont {Schmidgall}, \citenamefont {McGuire}, \citenamefont
  {Cobden} \emph {et~al.}}]{huang2017}%
  \BibitemOpen
  \bibfield  {author} {\bibinfo {author} {\bibfnamefont {B.}~\bibnamefont
  {Huang}}, \bibinfo {author} {\bibfnamefont {G.}~\bibnamefont {Clark}},
  \bibinfo {author} {\bibfnamefont {E.}~\bibnamefont {Navarro-Moratalla}},
  \bibinfo {author} {\bibfnamefont {D.~R.}\ \bibnamefont {Klein}}, \bibinfo
  {author} {\bibfnamefont {R.}~\bibnamefont {Cheng}}, \bibinfo {author}
  {\bibfnamefont {K.~L.}\ \bibnamefont {Seyler}}, \bibinfo {author}
  {\bibfnamefont {D.}~\bibnamefont {Zhong}}, \bibinfo {author} {\bibfnamefont
  {E.}~\bibnamefont {Schmidgall}}, \bibinfo {author} {\bibfnamefont {M.~A.}\
  \bibnamefont {McGuire}}, \bibinfo {author} {\bibfnamefont {D.~H.}\
  \bibnamefont {Cobden}}, \emph {et~al.},\ }\href@noop {} {\bibfield  {journal}
  {\bibinfo  {journal} {Nature}\ }\textbf {\bibinfo {volume} {546}},\ \bibinfo
  {pages} {270} (\bibinfo {year} {2017})}\BibitemShut {NoStop}%
\bibitem [{\citenamefont {Gao}\ \emph {et~al.}(2020)\citenamefont {Gao},
  \citenamefont {Zhou}, \citenamefont {Hong}, \citenamefont {Liang},
  \citenamefont {Song}, \citenamefont {Xu}, \citenamefont {Zhang},
  \citenamefont {Li}, \citenamefont {Wang},\ and\ \citenamefont
  {Dang}}]{gao2020}%
  \BibitemOpen
  \bibfield  {author} {\bibinfo {author} {\bibfnamefont {B.}~\bibnamefont
  {Gao}}, \bibinfo {author} {\bibfnamefont {T.}~\bibnamefont {Zhou}}, \bibinfo
  {author} {\bibfnamefont {A.-J.}\ \bibnamefont {Hong}}, \bibinfo {author}
  {\bibfnamefont {F.}~\bibnamefont {Liang}}, \bibinfo {author} {\bibfnamefont
  {G.}~\bibnamefont {Song}}, \bibinfo {author} {\bibfnamefont {Q.-Q.}\
  \bibnamefont {Xu}}, \bibinfo {author} {\bibfnamefont {J.}~\bibnamefont
  {Zhang}}, \bibinfo {author} {\bibfnamefont {G.-N.}\ \bibnamefont {Li}},
  \bibinfo {author} {\bibfnamefont {Y.}~\bibnamefont {Wang}},\ and\ \bibinfo
  {author} {\bibfnamefont {C.}~\bibnamefont {Dang}},\ }\href@noop {} {\bibfield
   {journal} {\bibinfo  {journal} {Appl. Phys. Express}\ }\textbf {\bibinfo
  {volume} {13}},\ \bibinfo {pages} {045001} (\bibinfo {year}
  {2020})}\BibitemShut {NoStop}%
\bibitem [{\citenamefont {Sheng}\ \emph {et~al.}(2020)\citenamefont {Sheng},
  \citenamefont {Zhu}, \citenamefont {Bai}, \citenamefont {Wu},\ and\
  \citenamefont {Wang}}]{sheng2020}%
  \BibitemOpen
  \bibfield  {author} {\bibinfo {author} {\bibfnamefont {H.}~\bibnamefont
  {Sheng}}, \bibinfo {author} {\bibfnamefont {Y.}~\bibnamefont {Zhu}}, \bibinfo
  {author} {\bibfnamefont {D.}~\bibnamefont {Bai}}, \bibinfo {author}
  {\bibfnamefont {X.}~\bibnamefont {Wu}},\ and\ \bibinfo {author}
  {\bibfnamefont {J.}~\bibnamefont {Wang}},\ }\href@noop {} {\bibfield
  {journal} {\bibinfo  {journal} {Nanotechnology}\ }\textbf {\bibinfo {volume}
  {31}},\ \bibinfo {pages} {315713} (\bibinfo {year} {2020})}\BibitemShut
  {NoStop}%
\bibitem [{\citenamefont {Hung}\ \emph {et~al.}(2017)\citenamefont {Hung},
  \citenamefont {Nugraha},\ and\ \citenamefont {Saito}}]{nguyen2017}%
  \BibitemOpen
  \bibfield  {author} {\bibinfo {author} {\bibfnamefont {N.~T.}\ \bibnamefont
  {Hung}}, \bibinfo {author} {\bibfnamefont {R.~T.}\ \bibnamefont {Nugraha},
  \bibfnamefont {A}},\ and\ \bibinfo {author} {\bibfnamefont {R.}~\bibnamefont
  {Saito}},\ }\href@noop {} {\bibfield  {journal} {\bibinfo  {journal} {App.
  Phys. Lett.}\ }\textbf {\bibinfo {volume} {111}},\ \bibinfo {pages} {092107}
  (\bibinfo {year} {2017})}\BibitemShut {NoStop}%
\bibitem [{\citenamefont {Cahaya}\ \emph {et~al.}(2015)\citenamefont {Cahaya},
  \citenamefont {Tretiakov},\ and\ \citenamefont {Bauer}}]{Cahaya}%
  \BibitemOpen
  \bibfield  {author} {\bibinfo {author} {\bibfnamefont {A.~B.}\ \bibnamefont
  {Cahaya}}, \bibinfo {author} {\bibfnamefont {O.~A.}\ \bibnamefont
  {Tretiakov}},\ and\ \bibinfo {author} {\bibfnamefont {G.~E.~W.}\ \bibnamefont
  {Bauer}},\ }\href@noop {} {\bibfield  {journal} {\bibinfo  {journal} {IEEE
  Trans. Magn.}\ }\textbf {\bibinfo {volume} {51}},\ \bibinfo {pages} {1}
  (\bibinfo {year} {2015})}\BibitemShut {NoStop}%
\bibitem [{\citenamefont {Giannozzi}\ \emph {et~al.}(2017)\citenamefont
  {Giannozzi}, \citenamefont {Andreussi}, \citenamefont {Brumme}, \citenamefont
  {Bunau}, \citenamefont {Nardelli}, \citenamefont {Calandra}, \citenamefont
  {Car}, \citenamefont {Cavazzoni}, \citenamefont {Ceresoli}, \citenamefont
  {Cococcioni}, \citenamefont {Colonna}, \citenamefont {Carnimeo},
  \citenamefont {Corso}, \citenamefont {de~Gironcoli}, \citenamefont {Delugas},
  \citenamefont {DiStasio}, \citenamefont {Ferretti}, \citenamefont {Floris},
  \citenamefont {Fratesi}, \citenamefont {Fugallo}, \citenamefont {Gebauer},
  \citenamefont {Gerstmann}, \citenamefont {Giustino}, \citenamefont {Gorni},
  \citenamefont {Jia}, \citenamefont {Kawamura}, \citenamefont {Ko},
  \citenamefont {Kokalj}, \citenamefont {K\"u{\c{c}}\"ukbenli}, \citenamefont
  {Lazzeri}, \citenamefont {Marsili}, \citenamefont {Marzari}, \citenamefont
  {Mauri}, \citenamefont {Nguyen}, \citenamefont {Nguyen}, \citenamefont {de-la
  Roza}, \citenamefont {Paulatto}, \citenamefont {Ponc{\'{e}}}, \citenamefont
  {Rocca}, \citenamefont {Sabatini}, \citenamefont {Santra}, \citenamefont
  {Schlipf}, \citenamefont {Seitsonen}, \citenamefont {Smogunov}, \citenamefont
  {Timrov}, \citenamefont {Thonhauser}, \citenamefont {Umari}, \citenamefont
  {Vast}, \citenamefont {Wu},\ and\ \citenamefont {Baroni}}]{QE}%
  \BibitemOpen
  \bibfield  {author} {\bibinfo {author} {\bibfnamefont {P.}~\bibnamefont
  {Giannozzi}}, \bibinfo {author} {\bibfnamefont {O.}~\bibnamefont
  {Andreussi}}, \bibinfo {author} {\bibfnamefont {T.}~\bibnamefont {Brumme}},
  \bibinfo {author} {\bibfnamefont {O.}~\bibnamefont {Bunau}}, \bibinfo
  {author} {\bibfnamefont {M.~B.}\ \bibnamefont {Nardelli}}, \bibinfo {author}
  {\bibfnamefont {M.}~\bibnamefont {Calandra}}, \bibinfo {author}
  {\bibfnamefont {R.}~\bibnamefont {Car}}, \bibinfo {author} {\bibfnamefont
  {C.}~\bibnamefont {Cavazzoni}}, \bibinfo {author} {\bibfnamefont
  {D.}~\bibnamefont {Ceresoli}}, \bibinfo {author} {\bibfnamefont
  {M.}~\bibnamefont {Cococcioni}}, \bibinfo {author} {\bibfnamefont
  {N.}~\bibnamefont {Colonna}}, \bibinfo {author} {\bibfnamefont
  {I.}~\bibnamefont {Carnimeo}}, \bibinfo {author} {\bibfnamefont {A.~D.}\
  \bibnamefont {Corso}}, \bibinfo {author} {\bibfnamefont {S.}~\bibnamefont
  {de~Gironcoli}}, \bibinfo {author} {\bibfnamefont {P.}~\bibnamefont
  {Delugas}}, \bibinfo {author} {\bibfnamefont {R.~A.}\ \bibnamefont
  {DiStasio}}, \bibinfo {author} {\bibfnamefont {A.}~\bibnamefont {Ferretti}},
  \bibinfo {author} {\bibfnamefont {A.}~\bibnamefont {Floris}}, \bibinfo
  {author} {\bibfnamefont {G.}~\bibnamefont {Fratesi}}, \bibinfo {author}
  {\bibfnamefont {G.}~\bibnamefont {Fugallo}}, \bibinfo {author} {\bibfnamefont
  {R.}~\bibnamefont {Gebauer}}, \bibinfo {author} {\bibfnamefont
  {U.}~\bibnamefont {Gerstmann}}, \bibinfo {author} {\bibfnamefont
  {F.}~\bibnamefont {Giustino}}, \bibinfo {author} {\bibfnamefont
  {T.}~\bibnamefont {Gorni}}, \bibinfo {author} {\bibfnamefont
  {J.}~\bibnamefont {Jia}}, \bibinfo {author} {\bibfnamefont {M.}~\bibnamefont
  {Kawamura}}, \bibinfo {author} {\bibfnamefont {H.-Y.}\ \bibnamefont {Ko}},
  \bibinfo {author} {\bibfnamefont {A.}~\bibnamefont {Kokalj}}, \bibinfo
  {author} {\bibfnamefont {E.}~\bibnamefont {K\"u{\c{c}}\"ukbenli}}, \bibinfo
  {author} {\bibfnamefont {M.}~\bibnamefont {Lazzeri}}, \bibinfo {author}
  {\bibfnamefont {M.}~\bibnamefont {Marsili}}, \bibinfo {author} {\bibfnamefont
  {N.}~\bibnamefont {Marzari}}, \bibinfo {author} {\bibfnamefont
  {F.}~\bibnamefont {Mauri}}, \bibinfo {author} {\bibfnamefont {N.~L.}\
  \bibnamefont {Nguyen}}, \bibinfo {author} {\bibfnamefont {H.-V.}\
  \bibnamefont {Nguyen}}, \bibinfo {author} {\bibfnamefont {A.~O.}\
  \bibnamefont {de-la Roza}}, \bibinfo {author} {\bibfnamefont
  {L.}~\bibnamefont {Paulatto}}, \bibinfo {author} {\bibfnamefont
  {S.}~\bibnamefont {Ponc{\'{e}}}}, \bibinfo {author} {\bibfnamefont
  {D.}~\bibnamefont {Rocca}}, \bibinfo {author} {\bibfnamefont
  {R.}~\bibnamefont {Sabatini}}, \bibinfo {author} {\bibfnamefont
  {B.}~\bibnamefont {Santra}}, \bibinfo {author} {\bibfnamefont
  {M.}~\bibnamefont {Schlipf}}, \bibinfo {author} {\bibfnamefont {A.~P.}\
  \bibnamefont {Seitsonen}}, \bibinfo {author} {\bibfnamefont {A.}~\bibnamefont
  {Smogunov}}, \bibinfo {author} {\bibfnamefont {I.}~\bibnamefont {Timrov}},
  \bibinfo {author} {\bibfnamefont {T.}~\bibnamefont {Thonhauser}}, \bibinfo
  {author} {\bibfnamefont {P.}~\bibnamefont {Umari}}, \bibinfo {author}
  {\bibfnamefont {N.}~\bibnamefont {Vast}}, \bibinfo {author} {\bibfnamefont
  {X.}~\bibnamefont {Wu}},\ and\ \bibinfo {author} {\bibfnamefont
  {S.}~\bibnamefont {Baroni}},\ }\href@noop {} {\bibfield  {journal} {\bibinfo
  {journal} {J. Phys. Condens. Matter}\ }\textbf {\bibinfo {volume} {29}},\
  \bibinfo {pages} {465901} (\bibinfo {year} {2017})}\BibitemShut {NoStop}%
\bibitem [{\citenamefont {Hamann}(2013)}]{hamann2013}%
  \BibitemOpen
  \bibfield  {author} {\bibinfo {author} {\bibfnamefont {D.~R.}\ \bibnamefont
  {Hamann}},\ }\href@noop {} {\bibfield  {journal} {\bibinfo  {journal} {Phys.
  Rev. B}\ }\textbf {\bibinfo {volume} {88}},\ \bibinfo {pages} {085117}
  (\bibinfo {year} {2013})}\BibitemShut {NoStop}%
\bibitem [{\citenamefont {Schlipf}\ and\ \citenamefont
  {Gygi}(2015)}]{schlipf2015}%
  \BibitemOpen
  \bibfield  {author} {\bibinfo {author} {\bibfnamefont {M.}~\bibnamefont
  {Schlipf}}\ and\ \bibinfo {author} {\bibfnamefont {F.}~\bibnamefont {Gygi}},\
  }\href@noop {} {\bibfield  {journal} {\bibinfo  {journal} {Comput. Phys.
  Commun.}\ }\textbf {\bibinfo {volume} {196}},\ \bibinfo {pages} {36}
  (\bibinfo {year} {2015})}\BibitemShut {NoStop}%
\bibitem [{\citenamefont {Perdew}\ \emph {et~al.}(1996)\citenamefont {Perdew},
  \citenamefont {Burke},\ and\ \citenamefont {Ernzerhof}}]{pbe1996}%
  \BibitemOpen
  \bibfield  {author} {\bibinfo {author} {\bibfnamefont {J.~P.}\ \bibnamefont
  {Perdew}}, \bibinfo {author} {\bibfnamefont {K.}~\bibnamefont {Burke}},\ and\
  \bibinfo {author} {\bibfnamefont {M.}~\bibnamefont {Ernzerhof}},\ }\href@noop
  {} {\bibfield  {journal} {\bibinfo  {journal} {Phys. Rev. Lett.}\ }\textbf
  {\bibinfo {volume} {77}},\ \bibinfo {pages} {3865} (\bibinfo {year}
  {1996})}\BibitemShut {NoStop}%
\bibitem [{\citenamefont {Mogulkoc}\ \emph {et~al.}(2020)\citenamefont
  {Mogulkoc}, \citenamefont {Modarresi},\ and\ \citenamefont
  {Rudenko}}]{Mogulkoc2020}%
  \BibitemOpen
  \bibfield  {author} {\bibinfo {author} {\bibfnamefont {A.}~\bibnamefont
  {Mogulkoc}}, \bibinfo {author} {\bibfnamefont {M.}~\bibnamefont
  {Modarresi}},\ and\ \bibinfo {author} {\bibfnamefont {A.~N.}\ \bibnamefont
  {Rudenko}},\ }\href@noop {} {\bibfield  {journal} {\bibinfo  {journal} {Phys.
  Rev. B}\ }\textbf {\bibinfo {volume} {102}},\ \bibinfo {pages} {024441}
  (\bibinfo {year} {2020})}\BibitemShut {NoStop}%
\bibitem [{\citenamefont {Mogulkoc}\ \emph {et~al.}()\citenamefont {Mogulkoc},
  \citenamefont {Modarresi},\ and\ \citenamefont {Rudenko}}]{Mogulkoc2021}%
  \BibitemOpen
  \bibfield  {author} {\bibinfo {author} {\bibfnamefont {A.}~\bibnamefont
  {Mogulkoc}}, \bibinfo {author} {\bibfnamefont {M.}~\bibnamefont
  {Modarresi}},\ and\ \bibinfo {author} {\bibfnamefont {A.}~\bibnamefont
  {Rudenko}},\ }\href@noop {} {\bibfield  {journal} {\bibinfo  {journal} {Phys.
  Rev. Applied}\ }\textbf {\bibinfo {volume} {15}},\ \bibinfo {pages}
  {064053} (\bibinfo {year} {2021})}\BibitemShut {NoStop}%
\bibitem [{SM()}]{SM}%
  \BibitemOpen
  \href@noop {} {}\bibinfo {note} {See supplementary material at ``URL". We
  give additional data for, (1) electronic band structures of all Cr$X$ without
  spin polarization, (2) phonon dispersion relations of all Cr$X$, (3)
  relaxation time model, and (4) thermoelectric transport coefficients of CrAs,
  CrSb, and CrBi.}\BibitemShut {Stop}%
\bibitem [{\citenamefont {Madsen}\ \emph {et~al.}(2018)\citenamefont {Madsen},
  \citenamefont {Carrete},\ and\ \citenamefont {Verstraete}}]{boltztrap2}%
  \BibitemOpen
  \bibfield  {author} {\bibinfo {author} {\bibfnamefont {G.~K.}\ \bibnamefont
  {Madsen}}, \bibinfo {author} {\bibfnamefont {J.}~\bibnamefont {Carrete}},\
  and\ \bibinfo {author} {\bibfnamefont {M.~J.}\ \bibnamefont {Verstraete}},\
  }\href@noop {} {\bibfield  {journal} {\bibinfo  {journal} {Comput. Phys.
  Commun.}\ }\textbf {\bibinfo {volume} {231}},\ \bibinfo {pages} {140}
  (\bibinfo {year} {2018})}\BibitemShut {NoStop}%
\bibitem [{\citenamefont {Ponc\'e}\ \emph {et~al.}(2016)\citenamefont
  {Ponc\'e}, \citenamefont {Margine}, \citenamefont {Verdi},\ and\
  \citenamefont {Giustino}}]{epw}%
  \BibitemOpen
  \bibfield  {author} {\bibinfo {author} {\bibfnamefont {S.}~\bibnamefont
  {Ponc\'e}}, \bibinfo {author} {\bibfnamefont {E.}~\bibnamefont {Margine}},
  \bibinfo {author} {\bibfnamefont {C.}~\bibnamefont {Verdi}},\ and\ \bibinfo
  {author} {\bibfnamefont {F.}~\bibnamefont {Giustino}},\ }\href@noop {}
  {\bibfield  {journal} {\bibinfo  {journal} {Comput. Phys. Commun.}\ }\textbf
  {\bibinfo {volume} {209}},\ \bibinfo {pages} {116} (\bibinfo {year}
  {2016})}\BibitemShut {NoStop}%
\bibitem [{\citenamefont {Togo}\ \emph {et~al.}(2015)\citenamefont {Togo},
  \citenamefont {Chaput},\ and\ \citenamefont {Tanaka}}]{phono3py1}%
  \BibitemOpen
  \bibfield  {author} {\bibinfo {author} {\bibfnamefont {A.}~\bibnamefont
  {Togo}}, \bibinfo {author} {\bibfnamefont {L.}~\bibnamefont {Chaput}},\ and\
  \bibinfo {author} {\bibfnamefont {I.}~\bibnamefont {Tanaka}},\ }\href@noop {}
  {\bibfield  {journal} {\bibinfo  {journal} {Phys. Rev. B}\ }\textbf {\bibinfo
  {volume} {91}},\ \bibinfo {pages} {094306} (\bibinfo {year}
  {2015})}\BibitemShut {NoStop}%
\bibitem [{\citenamefont {Mizokami}\ \emph {et~al.}(2018)\citenamefont
  {Mizokami}, \citenamefont {Togo},\ and\ \citenamefont {Tanaka}}]{phono3py2}%
  \BibitemOpen
  \bibfield  {author} {\bibinfo {author} {\bibfnamefont {K.}~\bibnamefont
  {Mizokami}}, \bibinfo {author} {\bibfnamefont {A.}~\bibnamefont {Togo}},\
  and\ \bibinfo {author} {\bibfnamefont {I.}~\bibnamefont {Tanaka}},\
  }\href@noop {} {\bibfield  {journal} {\bibinfo  {journal} {Phys. Rev. B}\
  }\textbf {\bibinfo {volume} {97}},\ \bibinfo {pages} {224306} (\bibinfo
  {year} {2018})}\BibitemShut {NoStop}%
\bibitem [{\citenamefont {Mao}\ \emph {et~al.}(2019)\citenamefont {Mao},
  \citenamefont {Zhu}, \citenamefont {Ding}, \citenamefont {Liu}, \citenamefont
  {Gamage}, \citenamefont {Chen},\ and\ \citenamefont {Ren}}]{mao2019high}%
  \BibitemOpen
  \bibfield  {author} {\bibinfo {author} {\bibfnamefont {J.}~\bibnamefont
  {Mao}}, \bibinfo {author} {\bibfnamefont {H.}~\bibnamefont {Zhu}}, \bibinfo
  {author} {\bibfnamefont {Z.}~\bibnamefont {Ding}}, \bibinfo {author}
  {\bibfnamefont {Z.}~\bibnamefont {Liu}}, \bibinfo {author} {\bibfnamefont
  {G.~A.}\ \bibnamefont {Gamage}}, \bibinfo {author} {\bibfnamefont
  {G.}~\bibnamefont {Chen}},\ and\ \bibinfo {author} {\bibfnamefont
  {Z.}~\bibnamefont {Ren}},\ }\href@noop {} {\bibfield  {journal} {\bibinfo
  {journal} {Science}\ }\textbf {\bibinfo {volume} {365}},\ \bibinfo {pages}
  {495} (\bibinfo {year} {2019})}\BibitemShut {NoStop}%
\bibitem [{\citenamefont {Hatami}\ \emph {et~al.}(2007)\citenamefont {Hatami},
  \citenamefont {Bauer}, \citenamefont {Zhang},\ and\ \citenamefont
  {Kelly}}]{PhysRevLett.99.066603}%
  \BibitemOpen
  \bibfield  {author} {\bibinfo {author} {\bibfnamefont {M.}~\bibnamefont
  {Hatami}}, \bibinfo {author} {\bibfnamefont {G.~E.~W.}\ \bibnamefont
  {Bauer}}, \bibinfo {author} {\bibfnamefont {Q.}~\bibnamefont {Zhang}},\ and\
  \bibinfo {author} {\bibfnamefont {P.~J.}\ \bibnamefont {Kelly}},\ }\href@noop
  {} {\bibfield  {journal} {\bibinfo  {journal} {Phys. Rev. Lett.}\ }\textbf
  {\bibinfo {volume} {99}},\ \bibinfo {pages} {066603} (\bibinfo {year}
  {2007})}\BibitemShut {NoStop}%
\bibitem [{\citenamefont {Hatami}\ \emph {et~al.}(2009)\citenamefont {Hatami},
  \citenamefont {Bauer}, \citenamefont {Zhang},\ and\ \citenamefont
  {Kelly}}]{PhysRevB.79.174426}%
  \BibitemOpen
  \bibfield  {author} {\bibinfo {author} {\bibfnamefont {M.}~\bibnamefont
  {Hatami}}, \bibinfo {author} {\bibfnamefont {G.~E.~W.}\ \bibnamefont
  {Bauer}}, \bibinfo {author} {\bibfnamefont {Q.}~\bibnamefont {Zhang}},\ and\
  \bibinfo {author} {\bibfnamefont {P.~J.}\ \bibnamefont {Kelly}},\ }\href@noop
  {} {\bibfield  {journal} {\bibinfo  {journal} {Phys. Rev. B}\ }\textbf
  {\bibinfo {volume} {79}},\ \bibinfo {pages} {174426} (\bibinfo {year}
  {2009})}\BibitemShut {NoStop}%
\end{thebibliography}

\begin{thebibliography}{5}%
\makeatletter
\providecommand \@ifxundefined [1]{%
 \@ifx{#1\undefined}
}%
\providecommand \@ifnum [1]{%
 \ifnum #1\expandafter \@firstoftwo
 \else \expandafter \@secondoftwo
 \fi
}%
\providecommand \@ifx [1]{%
 \ifx #1\expandafter \@firstoftwo
 \else \expandafter \@secondoftwo
 \fi
}%
\providecommand \natexlab [1]{#1}%
\providecommand \enquote  [1]{``#1''}%
\providecommand \bibnamefont  [1]{#1}%
\providecommand \bibfnamefont [1]{#1}%
\providecommand \citenamefont [1]{#1}%
\providecommand \href@noop [0]{\@secondoftwo}%
\providecommand \href [0]{\begingroup \@sanitize@url \@href}%
\providecommand \@href[1]{\@@startlink{#1}\@@href}%
\providecommand \@@href[1]{\endgroup#1\@@endlink}%
\providecommand \@sanitize@url [0]{\catcode `\\12\catcode `\$12\catcode
  `\&12\catcode `\#12\catcode `\^12\catcode `\_12\catcode `\%12\relax}%
\providecommand \@@startlink[1]{}%
\providecommand \@@endlink[0]{}%
\providecommand \url  [0]{\begingroup\@sanitize@url \@url }%
\providecommand \@url [1]{\endgroup\@href {#1}{\urlprefix }}%
\providecommand \urlprefix  [0]{URL }%
\providecommand \Eprint [0]{\href }%
\providecommand \doibase [0]{https://doi.org/}%
\providecommand \selectlanguage [0]{\@gobble}%
\providecommand \bibinfo  [0]{\@secondoftwo}%
\providecommand \bibfield  [0]{\@secondoftwo}%
\providecommand \translation [1]{[#1]}%
\providecommand \BibitemOpen [0]{}%
\providecommand \bibitemStop [0]{}%
\providecommand \bibitemNoStop [0]{.\EOS\space}%
\providecommand \EOS [0]{\spacefactor3000\relax}%
\providecommand \BibitemShut  [1]{\csname bibitem#1\endcsname}%
\let\auto@bib@innerbib\@empty
\bibitem [{\citenamefont {Giannozzi}\ \emph {et~al.}(2017)\citenamefont
  {Giannozzi}, \citenamefont {Andreussi}, \citenamefont {Brumme}, \citenamefont
  {Bunau}, \citenamefont {Nardelli}, \citenamefont {Calandra}, \citenamefont
  {Car}, \citenamefont {Cavazzoni}, \citenamefont {Ceresoli}, \citenamefont
  {Cococcioni}, \citenamefont {Colonna}, \citenamefont {Carnimeo},
  \citenamefont {Corso}, \citenamefont {de~Gironcoli}, \citenamefont {Delugas},
  \citenamefont {DiStasio}, \citenamefont {Ferretti}, \citenamefont {Floris},
  \citenamefont {Fratesi}, \citenamefont {Fugallo}, \citenamefont {Gebauer},
  \citenamefont {Gerstmann}, \citenamefont {Giustino}, \citenamefont {Gorni},
  \citenamefont {Jia}, \citenamefont {Kawamura}, \citenamefont {Ko},
  \citenamefont {Kokalj}, \citenamefont {K\"u{\c{c}}\"ukbenli}, \citenamefont
  {Lazzeri}, \citenamefont {Marsili}, \citenamefont {Marzari}, \citenamefont
  {Mauri}, \citenamefont {Nguyen}, \citenamefont {Nguyen}, \citenamefont {de-la
  Roza}, \citenamefont {Paulatto}, \citenamefont {Ponc{\'{e}}}, \citenamefont
  {Rocca}, \citenamefont {Sabatini}, \citenamefont {Santra}, \citenamefont
  {Schlipf}, \citenamefont {Seitsonen}, \citenamefont {Smogunov}, \citenamefont
  {Timrov}, \citenamefont {Thonhauser}, \citenamefont {Umari}, \citenamefont
  {Vast}, \citenamefont {Wu},\ and\ \citenamefont {Baroni}}]{QE}%
  \BibitemOpen
  \bibfield  {author} {\bibinfo {author} {\bibfnamefont {P.}~\bibnamefont
  {Giannozzi}}, \bibinfo {author} {\bibfnamefont {O.}~\bibnamefont
  {Andreussi}}, \bibinfo {author} {\bibfnamefont {T.}~\bibnamefont {Brumme}},
  \bibinfo {author} {\bibfnamefont {O.}~\bibnamefont {Bunau}}, \bibinfo
  {author} {\bibfnamefont {M.~B.}\ \bibnamefont {Nardelli}}, \bibinfo {author}
  {\bibfnamefont {M.}~\bibnamefont {Calandra}}, \bibinfo {author}
  {\bibfnamefont {R.}~\bibnamefont {Car}}, \bibinfo {author} {\bibfnamefont
  {C.}~\bibnamefont {Cavazzoni}}, \bibinfo {author} {\bibfnamefont
  {D.}~\bibnamefont {Ceresoli}}, \bibinfo {author} {\bibfnamefont
  {M.}~\bibnamefont {Cococcioni}}, \bibinfo {author} {\bibfnamefont
  {N.}~\bibnamefont {Colonna}}, \bibinfo {author} {\bibfnamefont
  {I.}~\bibnamefont {Carnimeo}}, \bibinfo {author} {\bibfnamefont {A.~D.}\
  \bibnamefont {Corso}}, \bibinfo {author} {\bibfnamefont {S.}~\bibnamefont
  {de~Gironcoli}}, \bibinfo {author} {\bibfnamefont {P.}~\bibnamefont
  {Delugas}}, \bibinfo {author} {\bibfnamefont {R.~A.}\ \bibnamefont
  {DiStasio}}, \bibinfo {author} {\bibfnamefont {A.}~\bibnamefont {Ferretti}},
  \bibinfo {author} {\bibfnamefont {A.}~\bibnamefont {Floris}}, \bibinfo
  {author} {\bibfnamefont {G.}~\bibnamefont {Fratesi}}, \bibinfo {author}
  {\bibfnamefont {G.}~\bibnamefont {Fugallo}}, \bibinfo {author} {\bibfnamefont
  {R.}~\bibnamefont {Gebauer}}, \bibinfo {author} {\bibfnamefont
  {U.}~\bibnamefont {Gerstmann}}, \bibinfo {author} {\bibfnamefont
  {F.}~\bibnamefont {Giustino}}, \bibinfo {author} {\bibfnamefont
  {T.}~\bibnamefont {Gorni}}, \bibinfo {author} {\bibfnamefont
  {J.}~\bibnamefont {Jia}}, \bibinfo {author} {\bibfnamefont {M.}~\bibnamefont
  {Kawamura}}, \bibinfo {author} {\bibfnamefont {H.-Y.}\ \bibnamefont {Ko}},
  \bibinfo {author} {\bibfnamefont {A.}~\bibnamefont {Kokalj}}, \bibinfo
  {author} {\bibfnamefont {E.}~\bibnamefont {K\"u{\c{c}}\"ukbenli}}, \bibinfo
  {author} {\bibfnamefont {M.}~\bibnamefont {Lazzeri}}, \bibinfo {author}
  {\bibfnamefont {M.}~\bibnamefont {Marsili}}, \bibinfo {author} {\bibfnamefont
  {N.}~\bibnamefont {Marzari}}, \bibinfo {author} {\bibfnamefont
  {F.}~\bibnamefont {Mauri}}, \bibinfo {author} {\bibfnamefont {N.~L.}\
  \bibnamefont {Nguyen}}, \bibinfo {author} {\bibfnamefont {H.-V.}\
  \bibnamefont {Nguyen}}, \bibinfo {author} {\bibfnamefont {A.~O.}\
  \bibnamefont {de-la Roza}}, \bibinfo {author} {\bibfnamefont
  {L.}~\bibnamefont {Paulatto}}, \bibinfo {author} {\bibfnamefont
  {S.}~\bibnamefont {Ponc{\'{e}}}}, \bibinfo {author} {\bibfnamefont
  {D.}~\bibnamefont {Rocca}}, \bibinfo {author} {\bibfnamefont
  {R.}~\bibnamefont {Sabatini}}, \bibinfo {author} {\bibfnamefont
  {B.}~\bibnamefont {Santra}}, \bibinfo {author} {\bibfnamefont
  {M.}~\bibnamefont {Schlipf}}, \bibinfo {author} {\bibfnamefont {A.~P.}\
  \bibnamefont {Seitsonen}}, \bibinfo {author} {\bibfnamefont {A.}~\bibnamefont
  {Smogunov}}, \bibinfo {author} {\bibfnamefont {I.}~\bibnamefont {Timrov}},
  \bibinfo {author} {\bibfnamefont {T.}~\bibnamefont {Thonhauser}}, \bibinfo
  {author} {\bibfnamefont {P.}~\bibnamefont {Umari}}, \bibinfo {author}
  {\bibfnamefont {N.}~\bibnamefont {Vast}}, \bibinfo {author} {\bibfnamefont
  {X.}~\bibnamefont {Wu}},\ and\ \bibinfo {author} {\bibfnamefont
  {S.}~\bibnamefont {Baroni}},\ }\href@noop {} {\bibfield  {journal} {\bibinfo
  {journal} {J. Phys. Condens. Matter}\ }\textbf {\bibinfo {volume} {29}},\
  \bibinfo {pages} {465901} (\bibinfo {year} {2017})}\BibitemShut {NoStop}%
\bibitem [{\citenamefont {Hamann}(2013)}]{hamann2013}%
  \BibitemOpen
  \bibfield  {author} {\bibinfo {author} {\bibfnamefont {D.~R.}\ \bibnamefont
  {Hamann}},\ }\bibfield  {title} {\bibinfo {title} {Optimized norm-conserving
  \text{Vanderbilt} pseudopotentials},\ }\href@noop {} {\bibfield  {journal}
  {\bibinfo  {journal} {Phys. Rev. B}\ }\textbf {\bibinfo {volume} {88}},\
  \bibinfo {pages} {085117} (\bibinfo {year} {2013})}\BibitemShut {NoStop}%
\bibitem [{\citenamefont {Schlipf}\ and\ \citenamefont
  {Gygi}(2015)}]{schlipf2015}%
  \BibitemOpen
  \bibfield  {author} {\bibinfo {author} {\bibfnamefont {M.}~\bibnamefont
  {Schlipf}}\ and\ \bibinfo {author} {\bibfnamefont {F.}~\bibnamefont {Gygi}},\
  }\bibfield  {title} {\bibinfo {title} {Optimization algorithm for the
  generation of \text{ONCV} pseudopotentials},\ }\href@noop {} {\bibfield
  {journal} {\bibinfo  {journal} {Comput. Phys. Commun.}\ }\textbf {\bibinfo
  {volume} {196}},\ \bibinfo {pages} {36} (\bibinfo {year} {2015})}\BibitemShut
  {NoStop}%
\bibitem [{\citenamefont {Perdew}\ \emph {et~al.}(1996)\citenamefont {Perdew},
  \citenamefont {Burke},\ and\ \citenamefont {Ernzerhof}}]{pbe1996}%
  \BibitemOpen
  \bibfield  {author} {\bibinfo {author} {\bibfnamefont {J.~P.}\ \bibnamefont
  {Perdew}}, \bibinfo {author} {\bibfnamefont {K.}~\bibnamefont {Burke}},\ and\
  \bibinfo {author} {\bibfnamefont {M.}~\bibnamefont {Ernzerhof}},\ }\bibfield
  {title} {\bibinfo {title} {Generalized gradient approximation made simple},\
  }\href@noop {} {\bibfield  {journal} {\bibinfo  {journal} {Phys. Rev. Lett.}\
  }\textbf {\bibinfo {volume} {77}},\ \bibinfo {pages} {3865} (\bibinfo {year}
  {1996})}\BibitemShut {NoStop}%
\bibitem [{\citenamefont {Ponc\'e}\ \emph {et~al.}(2016)\citenamefont
  {Ponc\'e}, \citenamefont {Margine}, \citenamefont {Verdi},\ and\
  \citenamefont {Giustino}}]{epw}%
  \BibitemOpen
  \bibfield  {author} {\bibinfo {author} {\bibfnamefont {S.}~\bibnamefont
  {Ponc\'e}}, \bibinfo {author} {\bibfnamefont {E.}~\bibnamefont {Margine}},
  \bibinfo {author} {\bibfnamefont {C.}~\bibnamefont {Verdi}},\ and\ \bibinfo
  {author} {\bibfnamefont {F.}~\bibnamefont {Giustino}},\ }\bibfield  {title}
  {\bibinfo {title} {\text{EPW}: Electron–phonon coupling, transport and
  superconducting properties using maximally localized wannier functions},\
  }\href@noop {} {\bibfield  {journal} {\bibinfo  {journal} {Comput. Phys.
  Commun.}\ }\textbf {\bibinfo {volume} {209}},\ \bibinfo {pages} {116}
  (\bibinfo {year} {2016})}\BibitemShut {NoStop}%
\bibitem [{\citenamefont {Madsen}\ \emph {et~al.}(2018)\citenamefont {Madsen},
  \citenamefont {Carrete},\ and\ \citenamefont {Verstraete}}]{boltztrap2}%
  \BibitemOpen
  \bibfield  {author} {\bibinfo {author} {\bibfnamefont {G.~K.}\ \bibnamefont
  {Madsen}}, \bibinfo {author} {\bibfnamefont {J.}~\bibnamefont {Carrete}},\
  and\ \bibinfo {author} {\bibfnamefont {M.~J.}\ \bibnamefont {Verstraete}},\
  }\bibfield  {title} {\bibinfo {title} {\text{BoltzTraP2}, a program for
  interpolating band structures and calculating semi-classical transport
  coefficients},\ }\href@noop {} {\bibfield  {journal} {\bibinfo  {journal}
  {Comput. Phys. Commun.}\ }\textbf {\bibinfo {volume} {231}},\ \bibinfo
  {pages} {140} (\bibinfo {year} {2018})}\BibitemShut {NoStop}%
\end{thebibliography}
%
\end{document}